\documentclass[twocolumn]{aastex631}

\usepackage{amsmath}

\shorttitle{AASTeX v6.3.1 Sample article}

\graphicspath{{./}{figures/}}

\begin{document}

\title{The Critical Core Mass of Rotating Planets}

\correspondingauthor{Cong Yu}
\email{yucong@mail.sysu.edu.cn}

\author{Wei Zhong}
\affiliation{School of Physics and Astronomy, Sun Yat-Sen University, Zhuhai, 519082, People's Republic of China}
\affiliation{State Key Laboratory of Lunar and Planetary Sciences, Macau University of Science and Technology, Macau, People's Republic of China}
\affiliation{CSST Science Center for the Guangdong-Hong Kong-Macau Greater Bay Area, Zhuhai, 519082, People's Republic of China}

\author{Cong Yu}
\affiliation{School of Physics and Astronomy, Sun Yat-Sen University, Zhuhai, 519082, People's Republic of China}
\affiliation{State Key Laboratory of Lunar and Planetary Sciences, Macau University of Science and Technology, Macau, People's Republic of China}
\affiliation{CSST Science Center for the Guangdong-Hong Kong-Macau Greater Bay Area, Zhuhai, 519082, People's Republic of China}

\begin{abstract}

The gravitational harmonics measured from Juno and Cassini spacecrafts help us to specify the internal structure and chemical elements of Jupiter and Saturn, respectively. However, we still do not know much about the impact of rotation on the planetary internal structure as well as their formation. The centrifugal force induced by rotation deforms the planetary shape and partially counteracts the gravitational force. Thus, rotation will affect the critical core mass of the exoplanet. Once the atmospheric mass becomes comparable to the critical core mass, the planet will enter the runaway accretion phase and becomes a gas giant. We have confirmed that the critical core masses of rotating planets depend on the stiffness of the polytrope, the outer boundary conditions, and the thickness of the isothermal layer. The critical core mass with Bondi boundary condition is determined by the surface properties. The critical core mass of a rotating planet will increase with the core gravity (i.e., the innermost density). For the Hill boundary condition, the soft polytrope shares the same properties as planets with Bondi boundary condition. Since the total mass for planets with Hill boundary condition increases with the decrease of the polytropic index, higher core gravity is required for rotating planets. As a result, the critical core mass in the stiff Hill model sharply increases. The rotation effects become more important when the radiative and convective regions coexist. 
Besides,  the critical core mass of planets with Hill (Bondi) boundary increases noticeably as the radiative layer becomes thinner (thicker). 

\end{abstract}


\section{Introduction} \label{sec:intro}

The core-nucleated instability is one of the mechanisms of the planet formation, which involves three different phases \citep{1986Bodenheimer,1996Icar..124...62P,2010PhT....63l..63A,2013Piso}. In the first phase, the rocky core accretes the planetesimal rapidly. In the second phase, the dust will be depleted within the neighboring region and the atmosphere envelope will grow gradually, which is regulated by the Kelvin-Helmholtz (KH) contraction. In the third phase, once the atmospheric envelope mass reaches the critical core mass, the planet will enter the runaway gas accretion phase and get inflated to be a gas giant. Note that when the runaway accretion occurs, the hydrostatic equilibrium of the envelope is difficult to preserve and the self-gravity of the gas envelope cannot be ignored \citep{2006ApJ...648..666R}.

The existence of the critical core mass suggests that when the envelope becomes massive enough and self-gravity of the envelope dominates, the envelope would collapse \citep{2019MNRAS.490.3144B},  the subsequent accretion would take place rapidly on a dynamical timescale. Before the runaway accretion, the accretion rate increases continuously with the core mass and the cooling of the envelope \citep{2013ApJ...765...33K}. Once the planet enters the runaway accretion, the accretion rate increases sharply. If the envelope mass approximately equals the core \citep{1986Bodenheimer} or the structure of the gaseous envelope is under the thermal and hydrostatic equilibrium \citep{2013ApJ...765...33K}, the maximum core mass would be ensured.

Many factors affect the critical core mass. \cite{2015ApJ...811...41L} hold that the metallic elements within the planet can increase the opacity of the envelope. As a result, the critical core mass will increase since the onset timescale of runaway accretion is proportional to the opacity.  \cite{2013ApJ...765...33K} verified that the stiffness of ploytrope, the disk condition around the accreting planets,  will change the distribution of the mass and density, and then, determine the critical core mass. Entropy advection \citep{2020Ali} may be a candidate mechanism to increase the critical core mass since it can shift radiative-convective boundary (i.e., RCB) inward.  In addition, the pebble isolation \citep{2020ApJ...896..135C} and tidally-forced turbulence \citep{2017ApJ...850..198Y}, can also change the planet's RCB and the condition of the critical core mass.

The planet will evolve with strong rotation support \citep{2019MNRAS.487.2319B}, when the core mass exceeds the thermal mass. Besides, the planets are rotating as it revolves around the central star due to the star-planet tidal interaction, which will influence the interior structure \citep{2019MNRAS.488.2365B}. Thus, rotation is a dispensable factor when considering the formation of the planets. We focus on the rotation of planets. Previous studies assumed that the planet is spherically symmetric. Actually, rotation will deform the shape (i.e., an oblate ellipsoid) and weaken the effect of gravity  \citep{2009PhT....62i..52M}. According to the conclusion of \cite{2013ApJ...765...33K}, we believe that rotation will affect the planet’s critical core mass. 

Currently, the internal structure of a rotating planet is investigated by matching the results of simulations of various mechanisms to the gravitational spherical harmonics obtained by Junno \citep{2010GeoRL..37.1204K} and Cassini \citep{2019Sci...364.2965I} spacecrafts. In general, there is a gap between the typical rigid body model of rotating planets \citep{1978ppi..book.....Z,Guillot1995CEPAM,2005AREPS..33..493G,2020A&A...639A..10N}
and the observation data. The gravitational harmonics derived by wind velocity in the deep flow \citep{2019GeoRL..46..616G,2019Sci...364.2965I} and metallic dynamo \citep{2019MNRAS.488.5633K} in the rotating planet can match the data. In addition, the spin velocity will increase with the planetary evolution until it gets to the maximum break-up velocity in which the planet dissociates \citep{2020MNRAS.491L..34G}.

We adopt  a rotating model proposed by \cite{2002A&A...394..965Z} to explore the effect of rotation on planetary evolution through obtaining the conditions of critical core mass. Following \cite{2013ApJ...765...33K}, we assume rotating planets evolve with different boundaries, i.e., Bondi or Hill boundaries. As mentioned above, rotation can change shape and partially counteract the gravitational force, our numerical calculations show that the planetary density and mass will condense toward the center and then slow the accretion processes (inhibit the runaway accretion ). As a result, the planet's evolutionary timescale will be prolonged. The critical core mass will increase finally.

This paper is structured as follows. In Section \ref{sec:intro}, we describe the structure of the rotating polytropic planet. In Section \ref{result}, we examine the effect of rotation on critical core mass and characteristic variables.  Section \ref{sec:result_polytrope} shows the effect of rotation on a single-layer polytrope. In Section \ref{sec:result_isothermal}, we study the effect of rotation in a composite polytropic planet, i.e., planet with an inner convective layer and an outer radiative layer smoothly connected at the radiative-convective bounary (RCB).  In Section \ref{conclusion}, we summarize the role of rotation in planetary formation.

\section{The model and assumption}\label{sec:intro}

In this section, we take rotation into account. Rotation alters the hydrostatic equilibrium with the coefficient of centrifugal force and then changes the critical core mass. This section divides into two parts. We exhibit the impact of rotation on the hydrostatic structure in Section \ref{spin}. Section \ref{Polytropic} shows the polytropic model. 

\subsection{The Hydrostatic Structure With Rotation}\label{spin}

Generally, rotation is a two-dimension problem. However, \cite{1997A&A...321..465M,2009PhT....62i..52M}  proposed a one-dimension shellular rotation model with constant angular velocity in the equipotential surfaces. The structure parameters of a rotating planet are evaluated by the mean values on equipotential surfaces.  However, \cite{2002A&A...394..965Z} provided an isobaric (equivalent) sphere can eliminate the mean values of the equipotential surface.

A rotating planet is an oblate ellipsoid, in which the volume on an isobar \citep{2002A&A...394..965Z} can be specified as
\begin{equation}
V = \frac{4}{3}\pi r^3 = \frac{4}{3}\pi \left(2R_{\rm e}^2+R_{\rm p}^2\right),
\label{volume}
\end{equation}
with the equatorial and polar radius, i.e., $R_{\rm e}$ and $R_{\rm p}$. $r$ denotes the isobaric radius. Since the planet is equipotential $\Psi ={\rm const}$, the total potential at colatitude $\theta$  can be written as \cite{2009PhT....62i..52M}
\begin{equation}
 \frac{GM}{r}+\frac{1}{2 }\omega^2r^2\sin^2\theta = \frac{GM}{R_{\rm p}},
\label{equatorial}
\end{equation}
where, $\omega$ is constant angular velocity.
The equatorial and polar radius are shown as ${R_{\rm e}}={R_{\rm p}}/{\left(1-\eta \right)}$ and ${R_{\rm p}}= r\left(1-\eta\right)^{2/3}$ with $\eta = \omega^2r^3/2GM$, respectively. Given the angular velocity and total mass, we can obtain the value of colatitude. Following \cite{2009PhT....62i..52M}, the maximal angular velocity $\omega_{\rm crit}$ is determined by the critical equatorial radius $R_{\rm e,crit}$ \citep{2009PhT....62i..52M}, that is,  $\omega_{\rm crit}^2 = {GM}/{R_{\rm e,crit}^3}$ 
with $R_{\rm e,crit} = 1.5 R_{\rm p}$. When the planet rotates with the break-up velocity, it will disintegrate \citep{2009PhT....62i..52M}. Thus, the angular velocity should be less than the break-up velocity.

As mentioned above, \cite{2009PhT....62i..52M}  has simplified the two-dimensional rotation problem as a one-dimensional quasi-static model with a constant rotation angle velocity,  resulting in the average density of element volume between two isobars, i.e., 
\begin{equation}
 \bar{\rho } \equiv \rho \frac{\left ( 1-r^2\sin^2 \theta \omega \alpha_{\rm p}\right )\left \langle g_{\rm geff}^{-1}\right \rangle}{\left \langle g_{\rm geff}^{-1}\right \rangle-\left \langle g_{\rm geff}^{-1} r^2 \sin^2 \theta\right \rangle \omega \alpha_{\rm p}},
 \label{eq:average_density}
\end{equation}
where, $g_{\rm eff}$ is the effective gravitational acceleration, the parameter $ \alpha_{\rm p}={\rm d} \omega/{\rm d} \Psi$ with potential $\Psi$.
Following \cite{2002A&A...394..965Z}, the average density in Equation (\ref{eq:average_density}) equals $\bar{\rho}=\rho \left \langle f_{\rm d}\right \rangle$, in which $\left \langle f_{\rm d}\right \rangle$ is treated as the correction factor. Following \cite{2013Stellar}, the mass between two isobars will switch into \citep{2002A&A...394..965Z}
\begin{equation}
{\rm d} M = 4\pi r^{\rm 2} \bar{\rho}=4\pi r^{\rm 2} \rho \left \langle f_{\rm d}\right \rangle {\rm d} r.
\label{eq:dmdr}
\end{equation}

The equivalent/isobaric sphere structure of a rotating planet is shown in Figure 1 of  \cite{2002A&A...394..965Z}. With a constant angular velocity,  we can obtain the colatitude of total potential  $\theta$ from Equation (\ref{equatorial}). Planetary spin imports a centrifugal acceleration ${a_{\rm n}} = \omega^{2} r \sin \theta$ that is perpendicular to the axis of rotation. Note that the planet's gravitational acceleration $g_{\rm r} = GM/r^{\rm 2}$ ($G$ is the gravitational constant) points to the geometric center of the equivalent sphere.  Therefore, there is a angle $\alpha_{\rm eff}$ between the effective gravitational and gravitational accelerations, and it satisfies 
 \begin{equation}
 \cos \alpha_{\rm eff} = \frac{g_{\rm r}-a_{\rm n}\cos \theta}{g_{\rm eff}} = \frac{g_{\rm r}^{2}+g_{\rm eff}^{2}-a_{\rm n}^{2}}{2g_{\rm r}g_{\rm eff}},
 \end{equation}
 with
 \begin{equation}
 g_{\rm eff} =  \left[\left(-g_{\rm r}+a_{\rm n} \sin \theta\right)^{2}+ \left(a_{\rm n} \cos \theta\right)^2 \right] ^{1/2}.
\label{g_eff}
\end{equation}
According to the structural diagram of \cite{2002A&A...394..965Z}, the relationship between the pressure and radius of the rotating planet is satisfied
\begin{equation}
\frac{{\rm d}P}{{\rm d}r} = -\bar{\rho}g =- \rho \left \langle f_{\rm d}\right \rangle g_{\rm eff} \cos \alpha_{\rm eff},
\label{eq:dp_dr}
\end{equation}
where, $g$ is the gravitational  acceleration  which points to the center. To simplify the relationship between pressure and radius, in Equation (\ref{eq:dp_dr}), \cite{2002A&A...394..965Z} defined the coefficient of centrifuge force
\begin{equation}
  f_{\rm p} \equiv  \frac{1}{2} \frac{g_{\rm r}^{2}+g_{\rm eff}^{2}-a_{\rm n}^{2}}{g_{\rm r}^{2}}.
\label{f_p}
\end{equation}
Thus, the pressure profile will change as follow
\begin{equation}
\frac{{\rm d}P}{{\rm d}r} = - \rho \left \langle f_{\rm d}\right \rangle g_{\rm r} f_{\rm p} = - \frac{GM}{r^{2}}\rho \left \langle f_{\rm d}\right \rangle  f_{\rm p}.
\label{eq:dp_dr_new}
\end{equation}
A planet spins at a constant angular velocity, the parameter $\alpha_{\rm p}={{\rm d} \omega}/{{\rm d} \Psi}=1$, the coefficient of the density $ \left \langle f_{\rm d}\right \rangle $ sets to $1$ \citep{2002A&A...394..965Z}.

In a rotating planet, the solid core is highly concentrated onto the center and immersed in the gaseous envelope \citep{1992ApJ...385..445F,2013ApJ...765...33K}. As the gas component is independent of the solid, they can be compatible with the different equations of state  \citep{1992ApJ...385..445F}. Their density and pressure satisfy $\rho = \rho_{\rm solid}+\rho_{\rm gas}$ and $P=P_{\rm solid}+P_{\rm gas}$, in which the subscript $gas$ and $solid$ correspond to the gas and solid component, respectively. {\bf Following Equations (\ref{eq:dmdr}) and (\ref{eq:dp_dr_new}), the} solid mass is given by
\begin{equation}
M_{\rm solid} = \int_{0}^{r}4\pi r^2 \rho_{\rm solid} \left \langle f_{\rm d}\right \rangle d{\rm r},
\label{mass_core}
\end{equation}
and the gaseous mass is
\begin{equation}
M_{\rm gas} = \int_{0}^{r}4\pi r^2 \rho_{\rm gas}  \left \langle f_{\rm d}\right \rangle d{\rm r}.
\label{mass_gas}
\end{equation}
The hydrostatic equilibrium can be changed by rotation
\begin{equation}
\frac{d P_{\rm solid}}{d {\rm r}} = -\frac{GM}{r^2} \rho_{\rm solid}  \left \langle f_{\rm d}\right \rangle f_{\rm p},
\label{solid_dp}
\end{equation}
\begin{equation}
\frac{d P_{\rm gas}}{d {\rm r}} = -\frac{GM}{r^2} \rho_{\rm gas}  \left \langle f_{\rm d}\right \rangle f_{\rm p}.
\label{gas_dp}
\end{equation}

The structure should satisfy the boundary conditions, which are described in Sections \ref{sec:interior} and \ref{sec:outerior}, respectively.  

\subsubsection{The Internal Boundary}\label{sec:interior}
Core radius  determines the internal boundary, which can be listed as \citep{2013ApJ...765...33K}
\begin{equation}
R_{\rm core} = \left( \frac{3M_{\rm solid}}{4\pi \rho_{\rm solid}}\right)^{1/3},
\label{r_core}    
\end{equation}
with the constant solid density.  Both the solid and gaseous component contribute to the core mass, thus,  $M_{\rm core}=M_{\rm solid}+M_{\rm gas}\left(R_{\rm core}\right)$ \citep{2013ApJ...765...33K}. Core density satisfy $\rho_{\rm core} = \rho_{\rm solid}+M_{\rm gas}\left(R_{\rm core}\right)/\left(4\pi R_{\rm core}^3/3\right)$, in which $\rho_{\rm solid}\gg \rho_{\rm gas}$ \citep{2013ApJ...765...33K}. The solid mass dominates in the core center.

\subsubsection{The Outer Boundary}\label{sec:outerior}

According to the theory of core accretion, the planet grows in the proto-planetary disk and accretes the matter from the disk. The disk parameters would determine the outer boundary condition. The density and pressure at the top of the envelope smoothly connect disk density and pressure (i.e., $\rho_{\rm disk}$ and $P_{\rm disk}$, \citealt{2013Chiang}), which can be given by
\begin{equation}
\rho=\rho_{\rm disk}=7.6 \times 10^{-9} \mathrm{~g} \mathrm{~cm}^{-3} a^{-2.9}, 
\label{density_out}
\end{equation}
\begin{equation}
P = P_{\rm disk} = 373 \mathrm{~K} \,\, a^{-3/7},
\label{out_p}
\end{equation}
where, $a$ is semi-major radius. The outer radius $R_{\rm out}$ is determined by the Bondi radius
\begin{equation}
R_{\rm B} = GM_{\rm p}/c_{\rm s}^2,
\label{bondi_radius}
\end{equation}
or Hill radius 
\begin{equation}
R_{\rm H} = a \left[M_{\rm p}/3\left(M_{\rm p}+M_{\star}\right)\right]^{1/3}.
\label{hill_out}
\end{equation}
The total mass $M_{\rm p}=M_{\rm core}+M_{\rm gas}\left(r\right)$ with the mass of the center star $M_{\star}$ and  the sonic speed $c_{\rm s} = \left(\gamma_{\rm disk}P_{\rm disk}/\rho_{\rm disk}\right)^{1/2}$ in proto-planet disk.

\subsection{The Homology Relation With Rotation}\label{Polytropic}

The homology relation of characteristic variables in a rotating planet is defined as follows
\begin{equation}
U \equiv \frac{d \log M}{d \log r}=\frac{\rho}{M /\left(4 \pi r^{3}\right)},
\label{U}
\end{equation}
\begin{equation}
V \equiv-\frac{d \log P}{d \log r}=\frac{G M / r}{P / \rho}f_{\rm p},
\label{V}
\end{equation}
where, Equation (\ref{U}) is same as \cite{1939isss.book.....C,1962PThPS..22....1H,2013ApJ...765...33K}. When planets evolve without rotation, $f_{\rm p}$ is equal to 1.

We employ a polytropic equation of state to describe the relationship between pressure and density
\begin{equation}
P = K \rho^{1+1/n},
\label{p_poly}
\end{equation}
with the polytropic index of $n=1-5$. The adiabatic constant $K$ is determined by the disk pressure and density, which can be read as \citep{2013ApJ...765...33K}
\begin{equation}
K = {P_{\rm disk}}/{\rho_{\rm disk}^{1+1/n}}.
\end{equation}
Combined with Equation (\ref{U}-\ref{V}), the polytropic index will connect to the homology invariant and satisfies  \citep{2013Stellar,2013ApJ...765...33K}
\begin{equation}
\frac{n}{n+1}=\frac{d \log \rho / d \log r}{d \log P / d \log r}.
\label{poly_index}
\end{equation}
Simplifying Equation (\ref{f_p}), we can obtain $f_{\rm p}=3-2\left(1-\eta\right)^{-2/3}$. The  hydrostatic equilibrium will switch to 
\begin{equation}
\frac{d\log U}{d\log V} = -\frac{U+V{\rm n}/\left(\rm n+1\right)-3}{\left(\alpha+1\right)U+V/\left(\rm n+1\right)-\left(3\alpha+1\right)}
\label{dlnU_dlnV}
\end{equation}
with $\alpha = 4\eta /\left[9 \left(1-\eta\right)^{5/3}-6 \left(1-\eta\right)\right]>0$. In the non-rotating planet, $\alpha = 0$ .

These parameters can characterize the properties of rotating planets by the simple $\log U-\log V$ diagram.  In addition, the radical radius satisfies 
\begin{equation}
\begin{split}
d\log r =& \frac{d\log M}{U} =  -\frac{d\log P}{V} \\
= &\frac{d\log \left ( V/U\right )}{\left(\alpha+2\right)U+V-\left ( 3\alpha +4\right )}.
\end{split}
\label{dlogr}
\end{equation}
The quantity $U/V$ in the rotating state represents the ratio of mass to pressure between the two isobars, i.e., $\left | {\rm d}\log M/{\rm d}\log P \right |$, which is consistent with the non-rotating state of \cite{2013ApJ...765...33K}. In contrast with the non-rotating state, the centrifugal acceleration coefficient $f_{\rm p}$ introduced by spin will change the value of $U/V$, which is approximate as $U/V \propto \left ( r^4 P\right )/\left ( M^2 f_{\rm p}\right )$ determined by mass, radius, pressure, and the angular rotation velocity. 
Following Equation (\ref{dlogr}), there is a critical line
\begin{equation}
\left(\alpha+2\right)U+V-\left ( 3\alpha +4\right )=0.
\label{critical_line}
\end{equation}
As $\alpha=0$, Equation (\ref{critical_line}) is consistent with \cite{2013ApJ...765...33K}. When $\left(\alpha+2\right)U+V-\left ( 3\alpha +4\right )>0$, $U/V$ will increases from the exterior to the interior along the structure lines. Instead, $U/V$ will increase from the inner to the outer shell.  There are vertical and horizontal lines, defined from the differentials of $U$ and $V$,  shown as follows
\begin{equation}
    U+{\rm n}V/\left(\rm n+1\right)-3 = 0,
\label{vertical}
\end{equation}
\begin{equation}
    \left(\alpha+1\right)U+V/\left(\rm n+1\right)-\left(3\alpha+1\right) = 0.
\label{horizontal_line}
\end{equation}
For $n\geq 3$, an intersection exists between these three lines. Hence, $U$ and $V$ see the need for
\begin{equation}
U=\left \{\left [ 3\left ( \alpha +1\right )-2\right ] n -3\right \}/\left [ \left ( \alpha +1\right )n-1\right ],
\end{equation}
\begin{equation}
V=2\left ( n+1\right )/\left [ \left ( \alpha +1\right )n-1\right ].
\end{equation}

The outer boundary conditions, connecting to the homology invariants, are given by 
\begin{equation}
U_{\rm surf,B} = \gamma_{\rm disk}^{-3}\left(M_{\rm p}/M_{\rm 0}\right)^{2}
\label{B_condition_u}
\end{equation}
\begin{equation}
V_{\rm surf,B} = \gamma_{\rm disk}f_{\rm p,surf},
\label{B_condition_V}
\end{equation}
with the coefficient of centrifugal force at surface $f_{\rm p,surf}$ and the characteristic mass $M_{\rm 0}=\left[\left(1/4\pi G^3\right)\left(P_{\rm disk}^3/\rho_{\rm disk}^4\right)\right]$  for the Bondi boundaries. The Hill boundaries can be read as follows
\begin{equation}
U_{\rm surf,H} = 4\pi a^3 \rho_{\rm disk}/\left(3 M_{\star}\right),
\label{H_condition_u}
\end{equation}
\begin{equation}
V_{\rm surf,H} = \left(3M_{\star}/4\pi a^3\rho_{\rm disk}\right)^{1/3}\left(M_{\rm p}/M_{\rm 0}\right)^{2/3}f_{\rm p,surf}.
\label{H_condition_v}
\end{equation}

The density between the top of the core and the bottom of the envelope is discontinuous \citep{2013ApJ...765...33K}, in which core density will jump to a high level. Other parameters are continuous.  The jump condition \citep{2013ApJ...765...33K} connecting the U-V plane shows
\begin{equation}
U_{\rm 1 e} / U_{\rm 1 i}=V_{\rm 1 e} / V_{\rm 1i}=\rho_{\rm 1e} / \rho_{\rm 1i},
\label{jump_condition}
\end{equation}
where the subscripts ${\rm 1e}$ and ${\rm 1i}$ correspond to the bottom and base of the envelope, respectively.

\section{Results}\label{result}
Rotation may change critical core mass and the homologous relation.  In Section \ref{sec:result_polytrope}, we discuss the effect of rotation in a single model. Section \ref{sec:result_isothermal} lists the critical mass of a rotating planet in the composite polytrope.


\subsection{The Results For A Single Polytrope}\label{sec:result_polytrope}
 In Section  \ref{sec:m_poly}, we explore how rotation changes the critical core mass in a single polytrope.  In Section \ref{sec:h_poly}, the homology structure of a rotating planet shows the different features.


\subsubsection{The Critical Core Mass of A Single Ploytrope}\label{sec:m_poly}

\begin{figure*}
\centering
\includegraphics[width = 8.6cm]{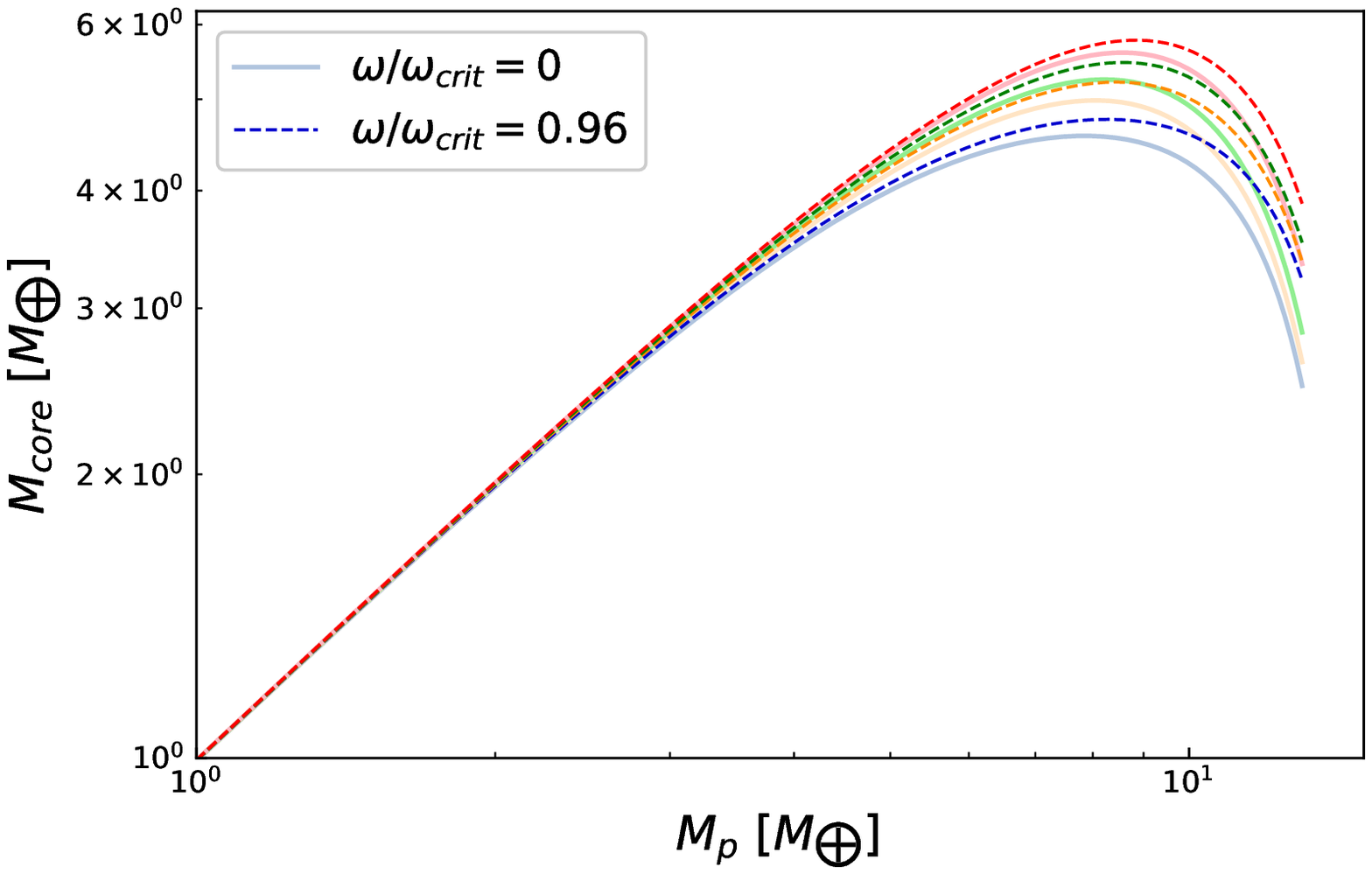}
\includegraphics[width = 8.6cm]{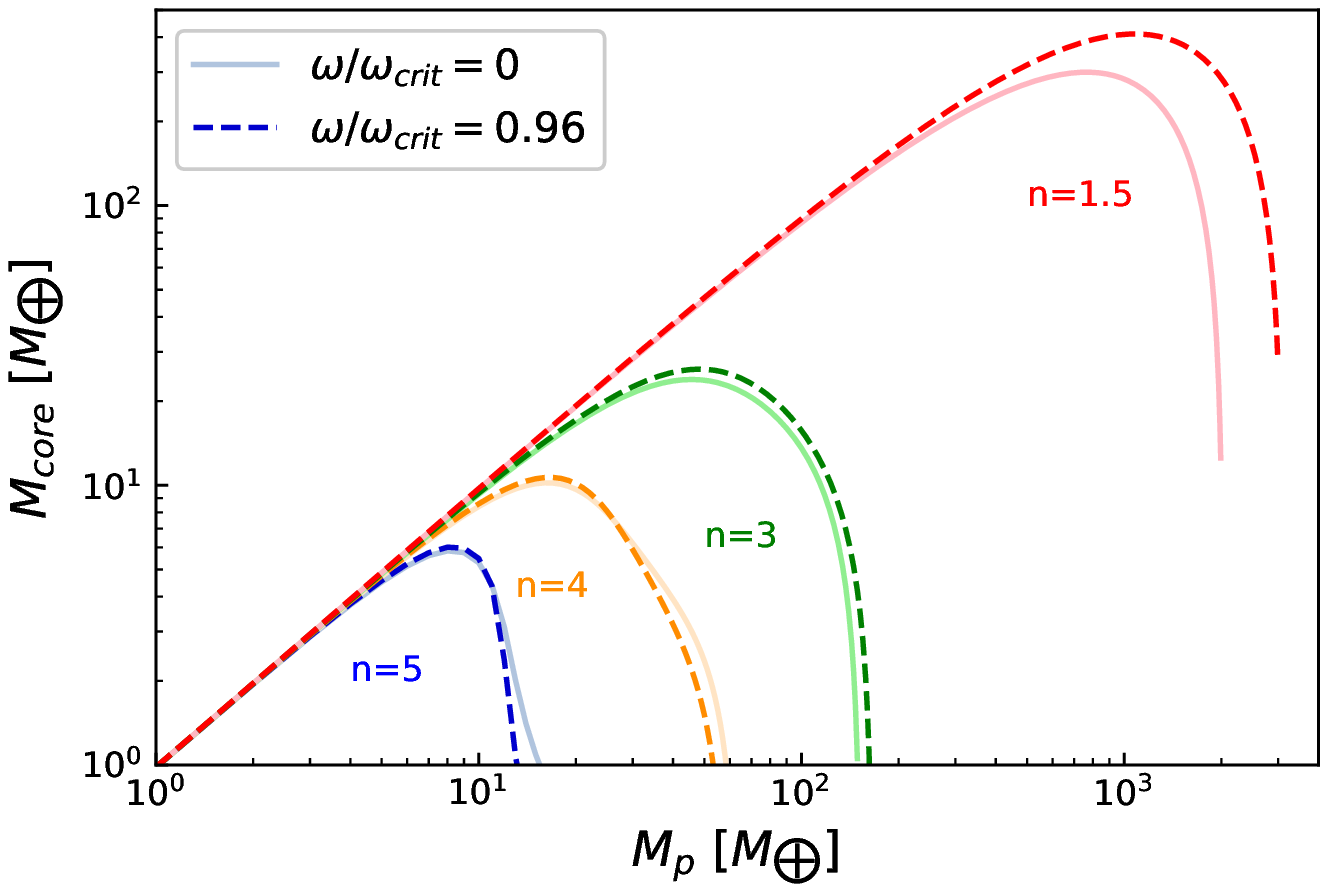}
\caption{The core mass is a function of the total mass in Bondi (left panel) and Hill models (right panel). The light blue, orange, green, and red lines correspond to the case of the non-rotating planet from $n = 1.5-5$, respectively. The dark dashed lines represent the rotating case. }
\label{critcal_mass}
\end{figure*}

\begin{deluxetable}{cc}
\tablenum{1}
\tablecaption{The input parameters\label{tab:input}}
\tablewidth{0pt}
\tablehead{
\colhead{Parameters} & \colhead{Value} 
}
\startdata
$M_{\star}$ & $1\,M_{\odot}$  \\
$\mu$ & 2.0  \\
$\rho_{\rm solid}$ & $5.5\,g\,cm^{-3}$  \\
$\gamma_{\rm disk}$ & 1  \\
$\omega/\omega_{\rm crit}$ & 0.96  \\
$a$ & 0.1 AU \\
$\rho_{\rm disk}$ & $6 \times 10^{-6}\,g\, cm^{-3}$  \\
$T_{\rm disk}$ & $1000\,K$  \\
\enddata
\end{deluxetable}

\begin{deluxetable*}{cccccccccc}
\tablenum{2}
\tablecaption{The critical parameters of a single polytrope \label{tab:critical}}
\tablewidth{0pt} 
\tablehead{
\colhead{n} & \colhead{$M_{\rm core}^{\rm crit}$} &  \colhead{$M_{\rm core,ro}^{\rm crit}$} & \colhead{$M_{\rm p}^{\rm crit} $}& \colhead{$M_{\rm p,ro}^{\rm crit} $}&\colhead{$U_{\rm 1e}^{\rm crit}$} &\colhead{$U_{\rm 1e,ro}^{\rm crit}$} & \colhead{$V_{\rm 1e}^{\rm crit}$} & \colhead{$V_{\rm 1e,ro}^{\rm crit}$} 
}
\startdata
&&&Bondi& &model&&\\
\hline
1.5 & 5.60 & 5.78  & 8.57 & 8.84& 0.00028 & 0.00030  &2.40 &2.41\\
3 &  5.24 & 5.47 & 8.21 & 8.58 &0.00613&0.00653&3.68 &3.70\\
4 & 4.98 & 5.21 & 8.04 & 8.45&0.0316&0.034&4.43&4.47\\
\hline
\hline
&&& Hill &&model&&&\\
\hline
1.5 & 299.68 & 410.53 & 761&1081 &0.0162&0.0221&2.32&2.31\\
3 &  23.89 & 26.00 & 46.00 &48.00 &0.177&0.207&3.30&3.31\\
4 & 10.20 & 10.66 & 16.0 & 17.0&0.327&0.386&4.02&3.98\\
\enddata
\end{deluxetable*}

To find the critical core mass, we first need to ensure the relationship between the core mass and the total mass. The suitable  solid mass determines the evolution of this core mass, which can be searched by solving the Equations  (\ref{U}) and (\ref{V}). The input parameters are listed in Table \ref{tab:input}. Figure  \ref{critcal_mass}  shows core mass evolves with the total mass in Bondi and Hill models, respectively. The characteristic variables of the critical models are listed in Table \ref{tab:critical}.

In the early evolution, the total mass is approximate to that of the core since the envelope mass was much lower than the core ($M_{\rm env} = M_{\rm p}-M_{\rm core}\ll M_{\rm core}$, \citealt{2006ApJ...648..666R,2013ApJ...765...33K}). We can integrate Equation (\ref{V}) with the polytropic relation from the outside to the inside. Finally, the density distribution satisfies 
\begin{equation}
\rho = \rho_{\rm disk} \left[1+\frac{V_{\rm surf} f_{\rm p}}{n+1}\left(\frac{R_{\rm out}}{r}-1\right)\right]^{n},
\label{density_profile}
\end{equation}
with the surface parameter $V_{\rm surf}= GM_{\rm core}\rho_{\rm disk}/R_{\rm out} P_{\rm disk} $. The density profile is  gentle relatively when $V_{\rm surf}f_{\rm p}/\left(n+1\right)<1$. As $V_{\rm surf}f_{\rm p}/\left(n+1\right)>1$, the density curve will become steeper. We can assume both the core mass and total mass for the non-rotating and rotating planets are constant. The density will decrease with the coefficient of centrifugal force ($f_{\rm p}$) induced by rotation.  However, in this situation, the core density (core mass) changes in a manner opposite to our assumption. To counteract this, we can obtain the same core mass by reducing the planetary mass. When the total mass is known, the core gravity rises sharply as a result of storing so much mass within the planet. Consequently, the core mass increases significantly.


As seen in Figure \ref{critcal_mass}, the critical core mass divides the accretion process into two parts. Core mass first increases with the total mass. The atmospheric mass is much lower than the core, so that core gravity determines the structure features. Once the core mass becomes comparable to the critical core mass, it will monotonically decrease.  Combined with Equation (\ref{U}) and Equation (\ref{density_profile}), we can derive the mass ratio between the envelope and core
\begin{equation}
\begin{split}
\frac{M_{\rm env}}{M_{\rm core}} = &\frac{4\pi  \rho_{\rm disk}R_{\rm out}^{3}}{M_{\rm core}} \\
&\times \int_{{R_{\rm core}}/R_{\rm out}}^{1} \xi^{3-n} \left [ \xi +\frac{f_{\rm p}V_{\rm surf}}{n+1}\left ( 1-\xi \right )\right ]^{\rm n}{\rm d}\log \xi,
\end{split}
\label{mass_distribution}    
\end{equation}
with $\xi = r/R_{\rm out}$. The atmospheric mass is different in Bondi and Hill models.  

In the Bondi model of  $V_{\rm surf}f_{\rm p}/\left(n+1\right)<1$, the density distribution is relatively uniform so that the atmospheric mass is approximate to the product of density and volume. As long as $R_{\rm out} \gg R_{\rm core}$, the mass ratio will change to ${M_{\rm env}}/{M_{\rm core}}\approx  {4 \pi \rho_{\rm disk}R_{\rm out}^3}/{3 M_{\rm core}} =\left(M_{\rm core}/M_{\rm 0}\right)^{\rm 2}/\gamma_{\rm disk}^{\rm 3}$  \citep{2013ApJ...765...33K}.  Many materials from the proto-planetary disk will be accreted onto the interior since both core gravity and mass will increase slightly by rotation. Note that the change in the critical core mass is insensitive to the polytropic index since the density for the Bondi model is flattest.

The Hill model covers three scenarios, $V_{\rm surf}f_{\rm p}/\left(n+1\right)<1$, $>1 $ or $\gg 1$. The value of $V_{\rm surf}f_{\rm p}/\left(n+1\right)$ affects the mass ratio by the density distribution. For $V_{\rm surf}f_{\rm p}/\left(n+1\right)<1$, it corresponds to the situation of $n<3$. Rotation will flatten the density distribution and then forces a mild increase in core gravity to keep the same planetary mass as the non-rotating planet.  Thus, the critical core mass will increase slightly. In the case of $V_{\rm surf}f_{\rm p}/\left(n+1\right)>1$, the innermost envelope determines the integral.  The innermost density is
\begin{equation}
\rho_{\rm 1e} = \rho_{\rm disk} \left(\frac{f_{\rm p}}{n+1}\right)^{n}\left(\frac{\rho_{\rm core}}{3\rho_{\rm disk}}\right)^{n/3}\left(\frac{M_{\rm core}}{M_{\rm 0}}\right)^{\rm 2n/3},
\label{density_innermost}
\end{equation}
when $R_{\rm out}/r \gg \min\left[1,V_{\rm surf}f_{\rm P}/\left(\rm n+1\right)\right]$ around core.
The mass ratio in the Hill model is independent of the outer radius and is determined by the properties of the innermost shell, which can be shown as follows  
\begin{equation}
\begin{split}
\frac{M_{\rm env}}{M_{\rm core}} \simeq & \frac{4\pi R_{\rm core}^{3}\rho_{\rm 1e}  }{3M_{\rm core}} \\
 =&\left(\frac{1}{3}\right)^{\rm n/3}\left(\frac{f_{\rm p}}{n+1}\right)^{n}\left(\frac{\rho_{\rm disk}}{\rho_{\rm core}}\right)^{\left(3-n\right)/3}\left(\frac{M_{\rm core}}{M_{\rm 0}}\right)^{2n/3}.
\label{mass_ratio_02}
\end{split}
\end{equation}
When $M_{\rm env} \sim M_{\rm core}$, the critical core mass can satisfy $M_{\rm core}^{\rm crit}/M_{\rm 0}\simeq \sqrt{3} \left[\left(n+1\right)/f_{\rm p}\right]^{3/2}  \left(\rho_{\rm disk}/\rho_{\rm core}\right)^{\left(n-3\right)/2n}$.  As analyzed above,  both the core density and the core mass will increase with the reduction in $f_{\rm p}$. Thus, the critical core mass may increase moderately. Under the same outer boundaries, we hold the atmospheric mass is reduced. 

When $V_{\rm surf}f_{\rm p}/\left(n+1\right)\gg 1$, the density will  increase rapidly, resulting in the largest mass concentration in the middle of the envelope. Therefore, the mass ratio will switch into
\begin{equation}
\frac{M_{\rm env}}{M_{\rm core}} \sim \left(\frac{f_{\rm p}}{n+1}\right)^{n} \left(\frac{\rho_{\rm disk}}{\left \langle \rho_{\rm \ast } \right \rangle}\right)^{\left(3-n\right)/3} \left(\frac{M_{\rm core}}{M_{\rm 0}}\right)^{2n/3},
\label{mass_ratio_hill_01}
\end{equation}
with the mean density $\left \langle \rho_{\rm \ast } \right \rangle = M_{\rm \ast}/\left(4\pi a^3/3\right)$.  Core gravity would significantly increase compared to other cases.  Besides, core mass in a rotating planet will noticeably increase because of $M_{\rm core}^{\rm crit} \propto \left[ \left({\rm n+1}\right)/f_{\rm p}\right]^{3/2}$.  

In summary, the density and mass distributions are relatively uniform in the soft Hill model ($n>3$ when $V_{\rm surf}f_{\rm p}/\left(n+1\right)< 1$ ). The critical core mass will be increased slightly by rotation. For $n=3$, the density distribution in the non-rotating model will increase moderately, which will be weakened by rotation.  Hence, the critical core mass for a rotating planet in this state will increase. The changing trend of density in the stiff Hill polytrope ($n<3$ when $V_{\rm surf}f_{\rm p}/\left(n+1\right)\gg 1$) is similar to the case of $n=3$, but  core mass increases noticeably. Thus, we hold that the critical core mass increases with the stiffness of the polytrope.

When the atmospheric mass approaches the core mass, the runaway accretion is triggered so that self-gravity becomes crucial. The planet will expel the envelope mass outward to adapt to the larger mass \citep{2013ApJ...765...33K}. Rotation can increase the core mass. We hold that the critical core mass of a single rotating polytropic model noticeably grows for the case with a Bondi radius or the softest Hill model ($n<3$).

 \begin{figure*}
    \centering
    \includegraphics[width = 8.6cm]{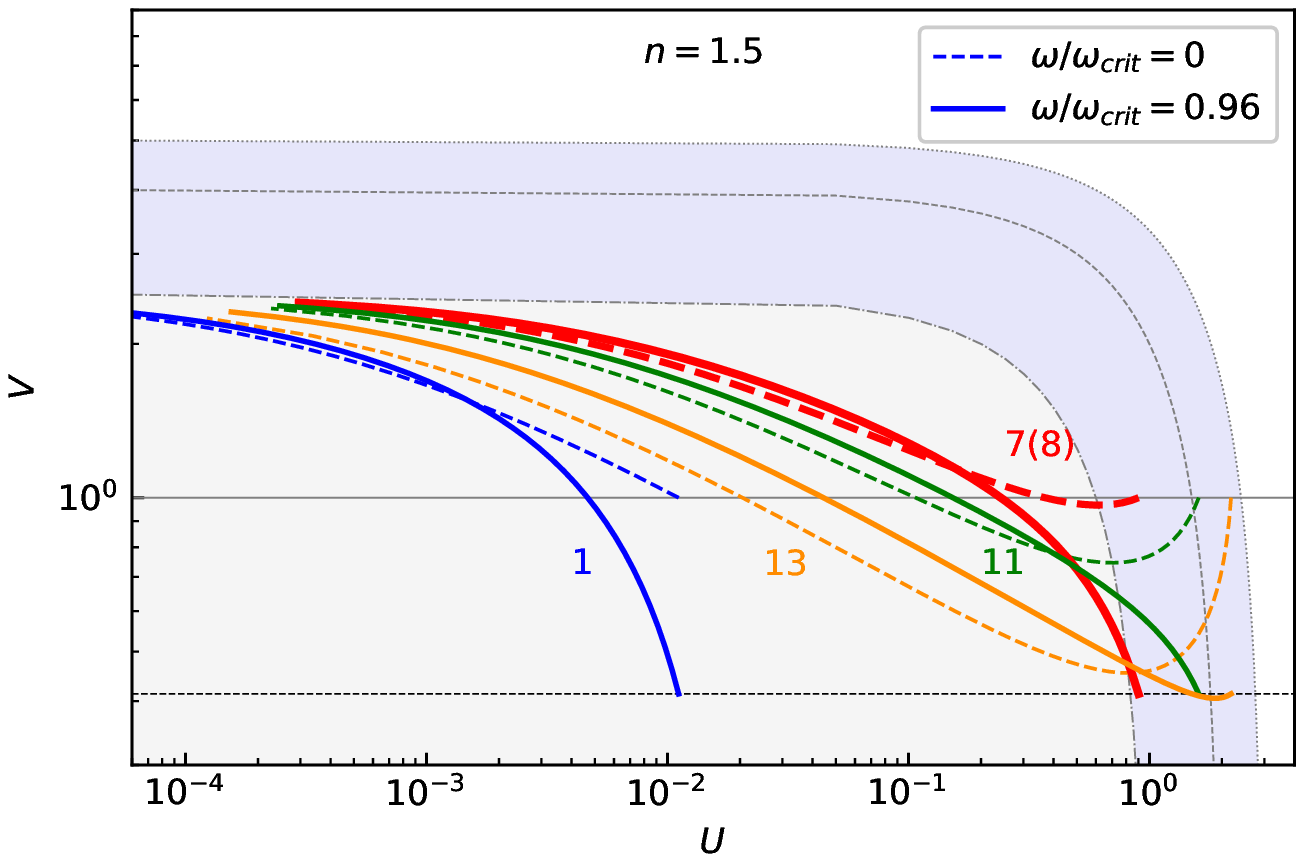}
    \includegraphics[width = 8.6cm]{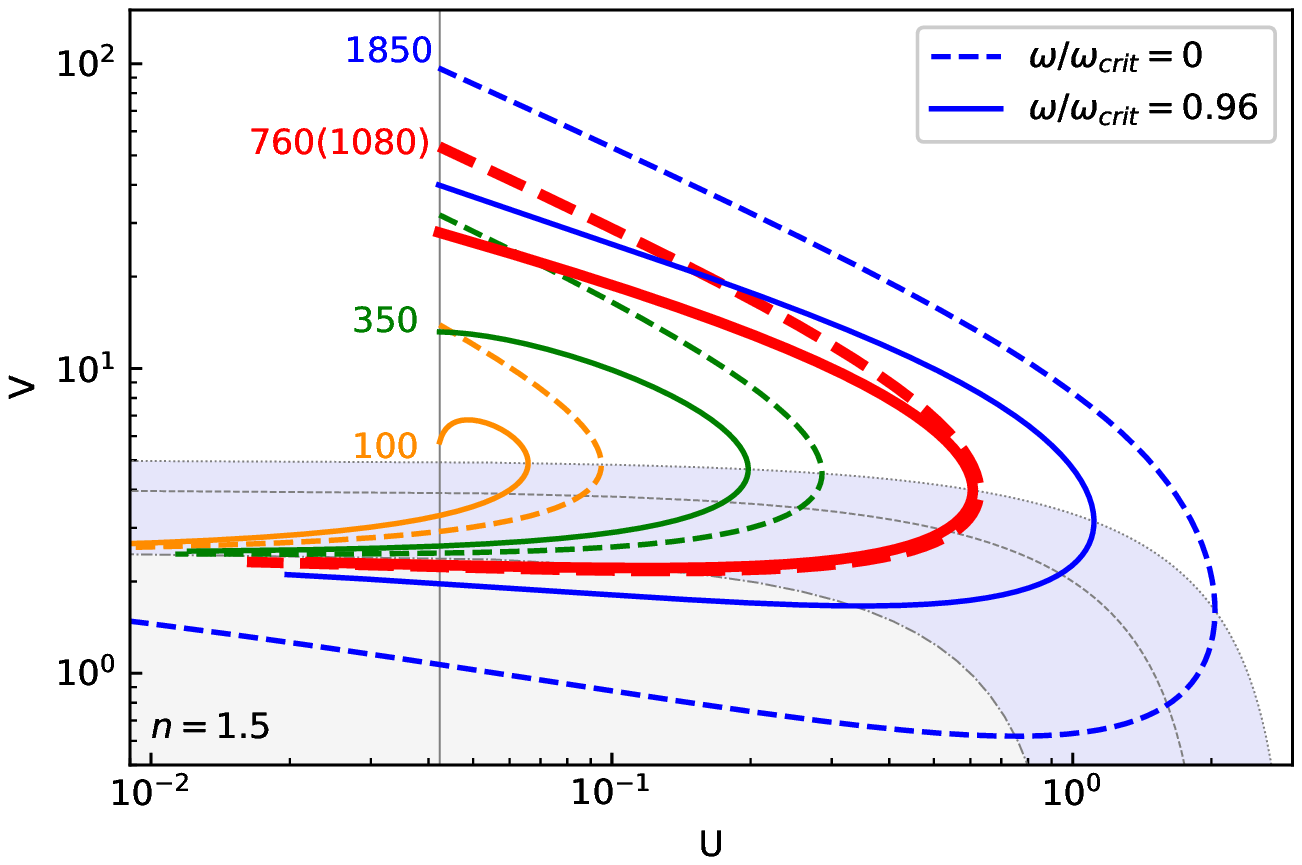}
    
    \includegraphics[width = 8.6cm]{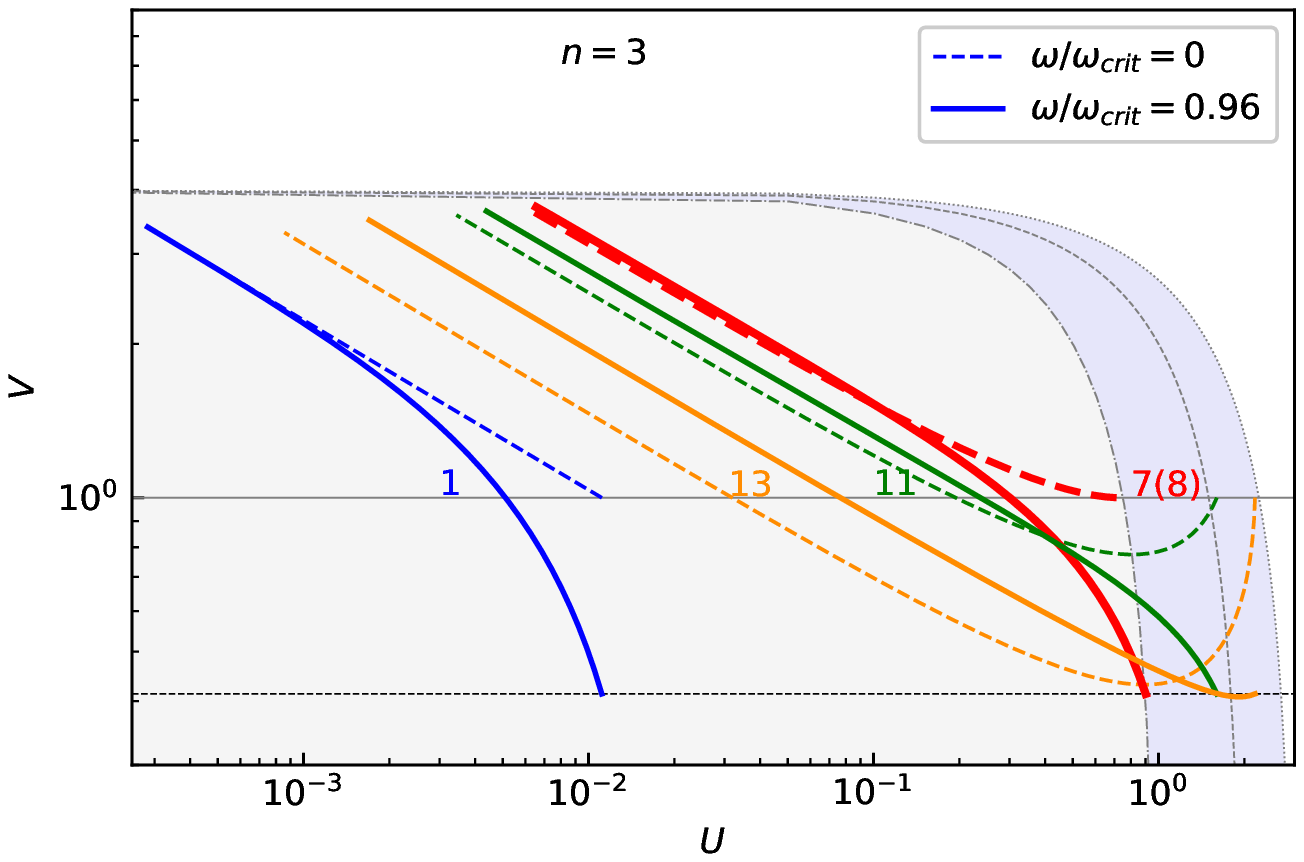}
    \includegraphics[width = 8.6cm]{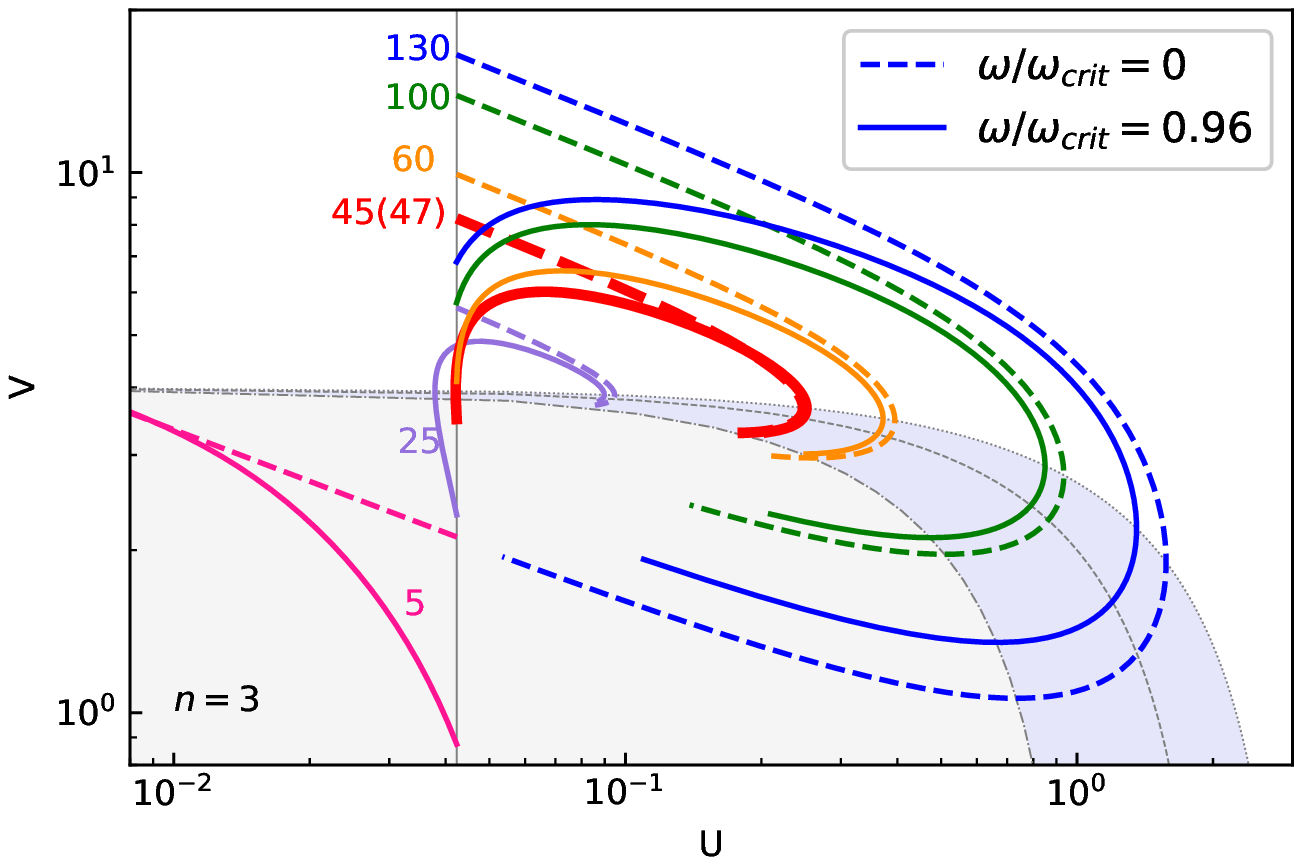}
    
     \includegraphics[width = 8.6cm]{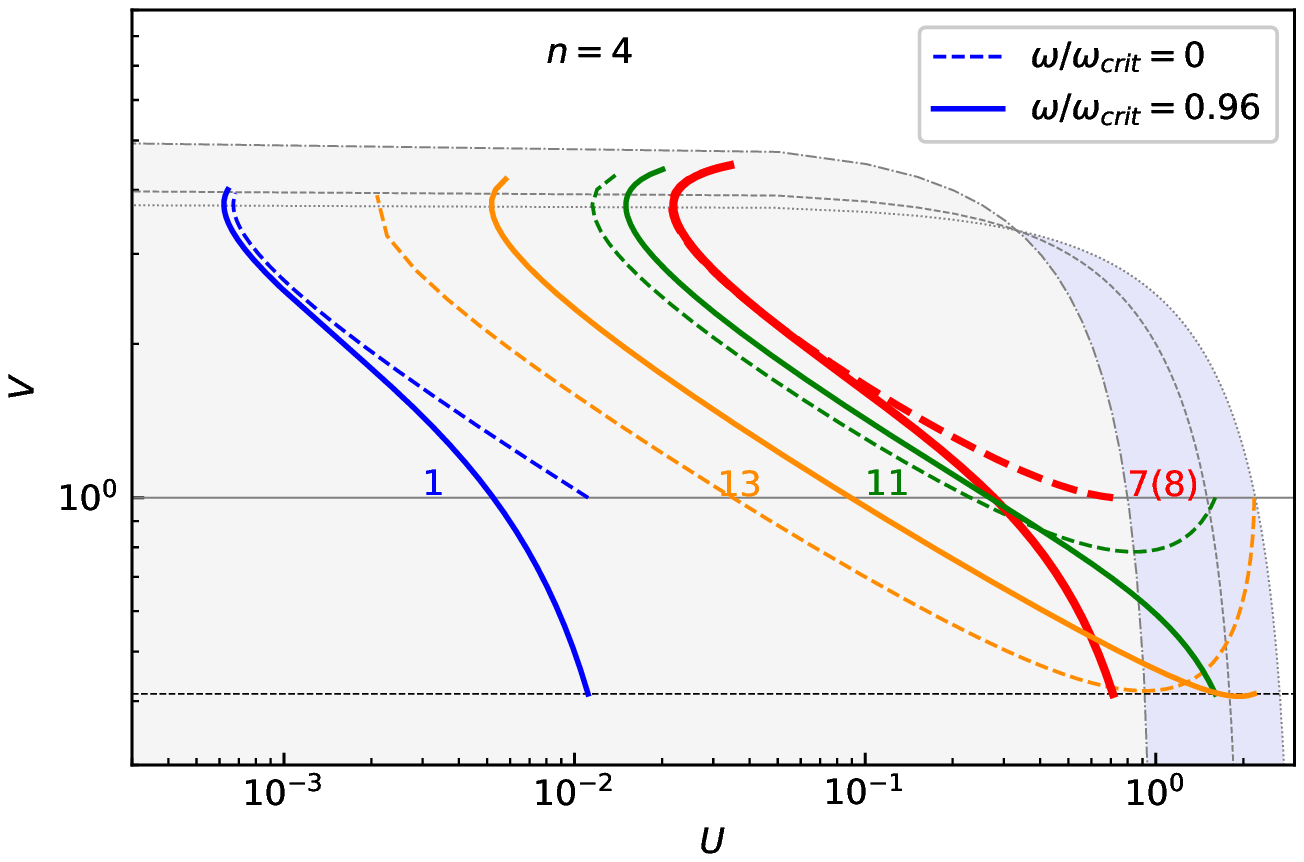}
    \includegraphics[width = 8.6cm]{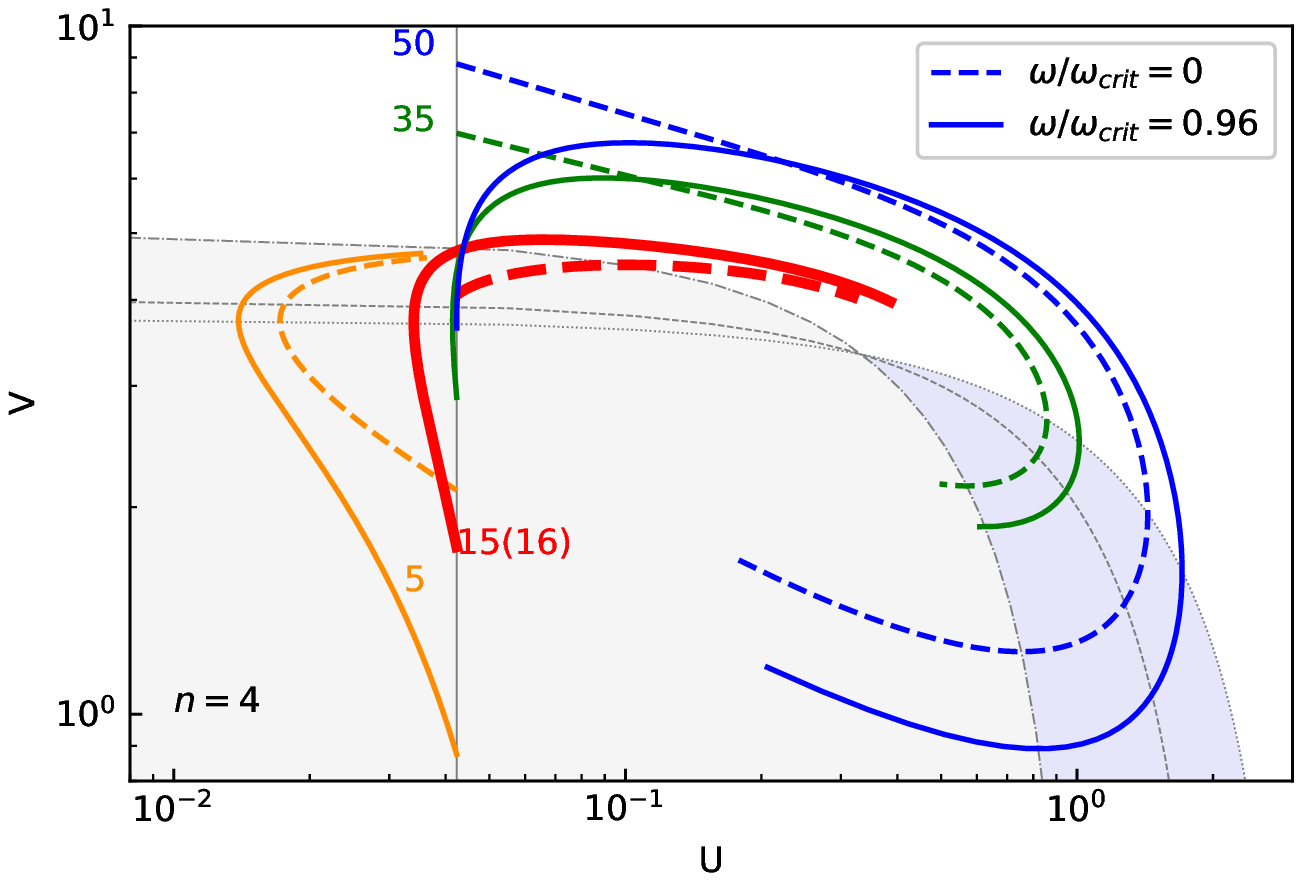}
    \caption{The structure line of a single polytropic model evolves with total mass. The left panels denote the Bondi models, and the right panels are the Hill models. From top to bottom, the polytropic index is specified as $N = 1.5-4$. Different colors represent different total masses ($M_{\rm \bigoplus}$). The values of the total masses are written by the corresponding colors. The numbers in parentheses are for the critical model for the rotating planets. Dotted lines represent the non-rotating case, and solid lines are for the rotating case. Red bold dashed and solid lines represent the critical models without and with spin, respectively. The solid gray and black lines represent the outer boundaries for the planets without and with rotation, respectively. Long dashed, broken, dotted lines denote the vertical, critical, horizontal lines for the non-rotating case, respectively. The coefficients of centrifugal force derived by the rotation are not constant, which increases inward.  
The critical and horizontal lines are close to the core. Besides, the efficiency of the rotation near the core is small. Thus, the differences of the characteristic lines between the planets without/with spin are much slight. }
    \label{fig:strucure_single}
\end{figure*}

\subsubsection{The Homolygy Relation of A Single Ploytrope}\label{sec:h_poly}

In this section, we mainly discuss the influence of rotation on the homology relation. As shown in Figure \ref{fig:strucure_single}, rotation will change the relationship between $U$ and $V$. Following Equation (\ref{density_profile}), the profiles of the characteristic variables $V$ and $U$ can be expressed as
\begin{equation}
V =  
 \left(n+1\right)/\left[\frac{r}{R_{\rm out}}\left(\frac{n+1}{V_{\rm surf}f_{\rm p}}-1\right)+1\right],
\label{v_new}
\end{equation}
\begin{equation}
U = 
\frac{4\pi r^3 \rho_{\rm disk}}{M_{\rm core}}\left[1+\frac{V_{\rm surf}f_{\rm p} }{n+1}\left(\frac{R_{\rm out}}{r}-1\right)\right]^{\rm n}.
\label{U_new}
\end{equation}
The parameter $V_{\rm surf}=GM_{\rm core}\rho_{\rm disk}/R_{\rm out}P_{\rm disk}$ will increase with the core mass. With the constant boundary conditions, the characteristic variables $U$ and $V$ will be affected by the coefficient of centrifugal force $f_{\rm p}$.

At the inner edge, the above relation will change into
\begin{equation}
\begin{split}
V_{\rm 1e} = &\frac{GM_{\rm core}f_{\rm p}\rho_{\rm 1e}}{P_{\rm 1e}R_{\rm core}} \\
= & f_{\rm p} \left( \frac{M_{\rm core}}{M_{\rm 0}}\right)^{\rm 2/3} \left(\frac{3\rho_{\rm disk}}{\rho_{\rm core}}\right)^{\rm \left(3-n\right)/3n} \left(\frac{3\rho_{\rm 1e}}{\rho_{\rm core}}\right)^{\rm -1/n},
\end{split}
\label{jump_condition_V}
\end{equation}
\begin{equation}
\begin{split}
U_{\rm 1e} \simeq & 3 \rho_{\rm 1e}/\rho_{\rm core}\\
\simeq & \left(\frac{f_{\rm p}}{n+1}\right)^{\rm n} \left(\frac{\rho_{\rm core}}{3\rho_{\rm disk}}\right)^{\rm {\left(n-3\right)}/{3}}\left(\frac{M_{\rm core}}{M_{\rm 0}}\right)^{\rm 2n/3} ,
\end{split}
\label{jump_condition_u}
\end{equation}
with $R_{\rm out}/r \gg \min \left[1, V_{\rm surf}f_{\rm p}/\left(n+1\right)\right]$. The jump condition satisfies
\begin{equation}
V_{\rm 1e} U_{\rm 1e}^{\rm 1/n} = f_{\rm p} \left( \frac{M_{\rm core}}{M_{\rm 0}}\right)^{\rm 2/3} \left(\frac{3\rho_{\rm disk}}{\rho_{\rm core}}\right)^{\rm \left(3-n\right)/3n}.
\label{tot_jump_condition}
\end{equation}

The structure line shows the unique characteristics of the Bondi and Hill models in Figure \ref{fig:strucure_single}.  All structure lines for the Bondi model are below the horizontal line. The horizontal line in a rotating planet changes to $V =\left[\left(3\alpha+1\right)-\left(\alpha+1\right)U\right]\left(n+1\right)$, in which $\alpha>0$. There are two different structure lines for the non-rotating or rotating case. First, the planet's mass is below the critical core mass,  implying core gravity is effective. The structure line will increase inwardly from the surface boundary until approaching the horizontal line \citep{2013ApJ...765...33K}. The centrifugal force, driven by rapid rotation, will weaken part of the gravitational force so that the value of $V_{\rm surf,ro}=\gamma_{\rm disk}f_{\rm p}$ at the outer boundary (black dashed line) would decrease sharply. While the exterior structural lines are much lower than that of the non-rotating planet, the interior will rise higher with the core mass.
Secondly, the planet's mass is above the critical core mass so that  the gravity of the envelope is effective. The structural line will pass through the outer boundary again and reach the minimum $V_{\rm min}$ at the horizontal line. In a  rotating planet, $V_{\rm min}$ is close to the boundary line. In summary, rotation changes the characteristic variables for the structural line at the same mass, resulting in an increase of critical core mass.

\begin{figure*}
\centering
\includegraphics[width = 8.5cm]{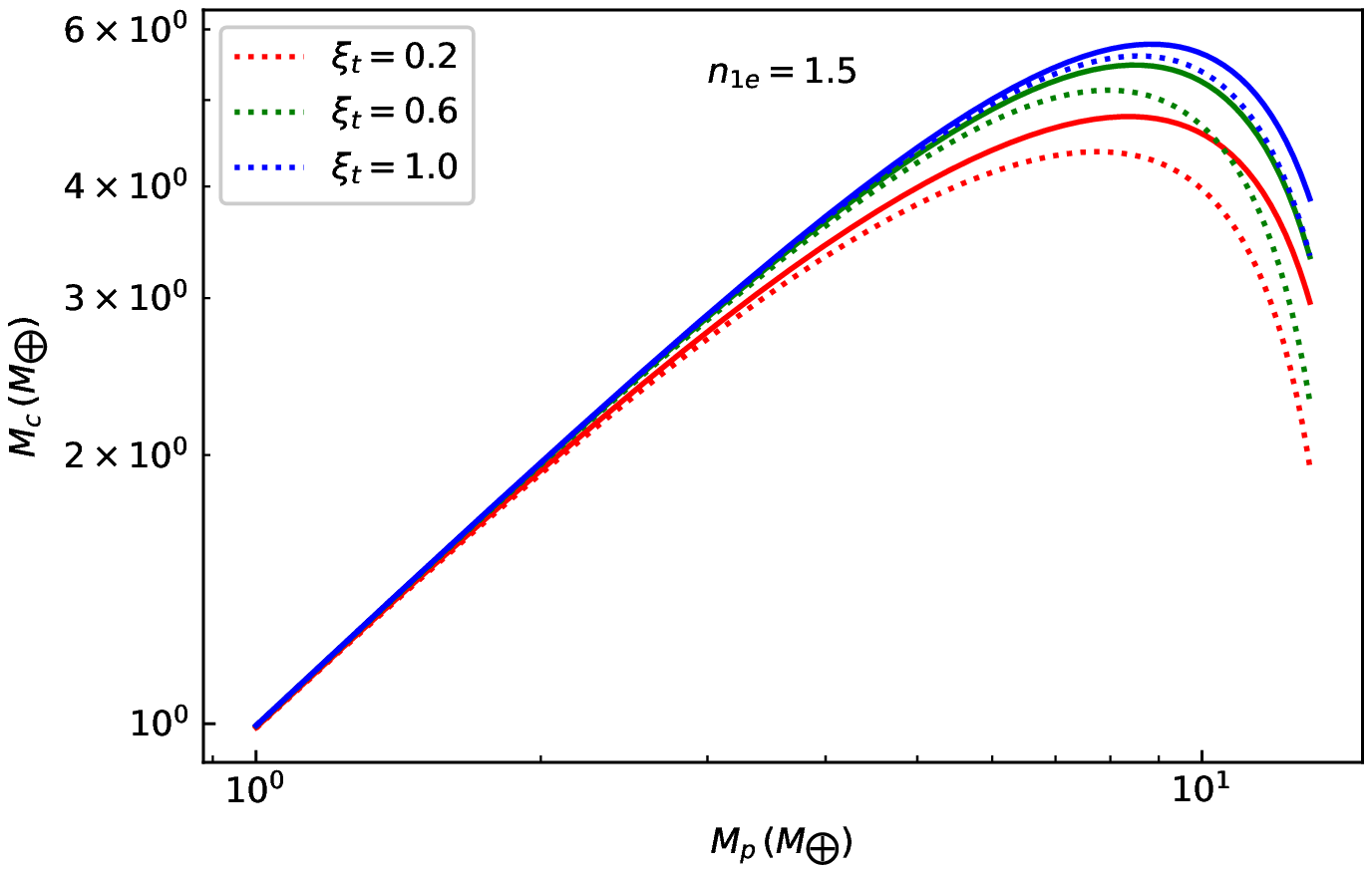}
\includegraphics[width = 8.5cm]{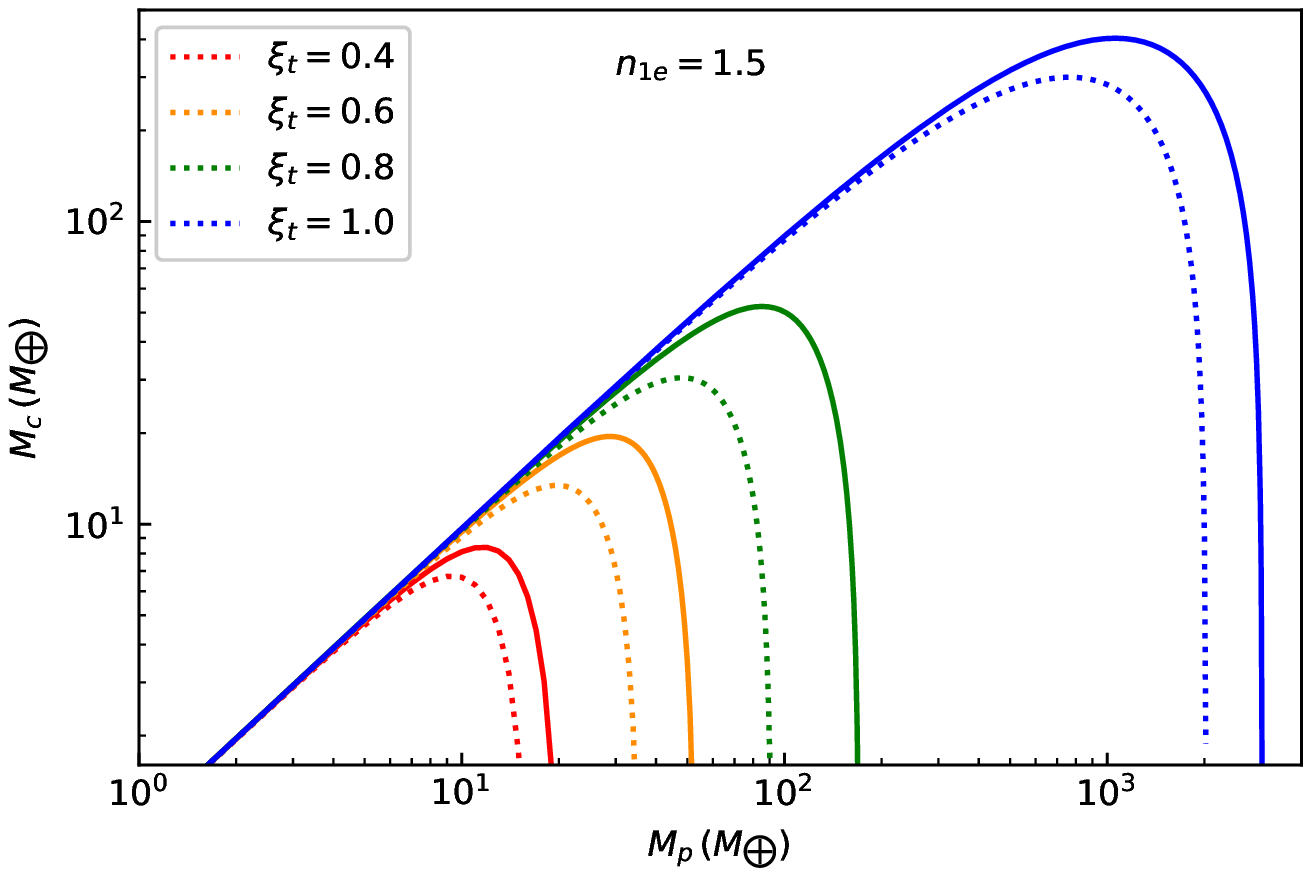}
    
\includegraphics[width = 8.5cm]{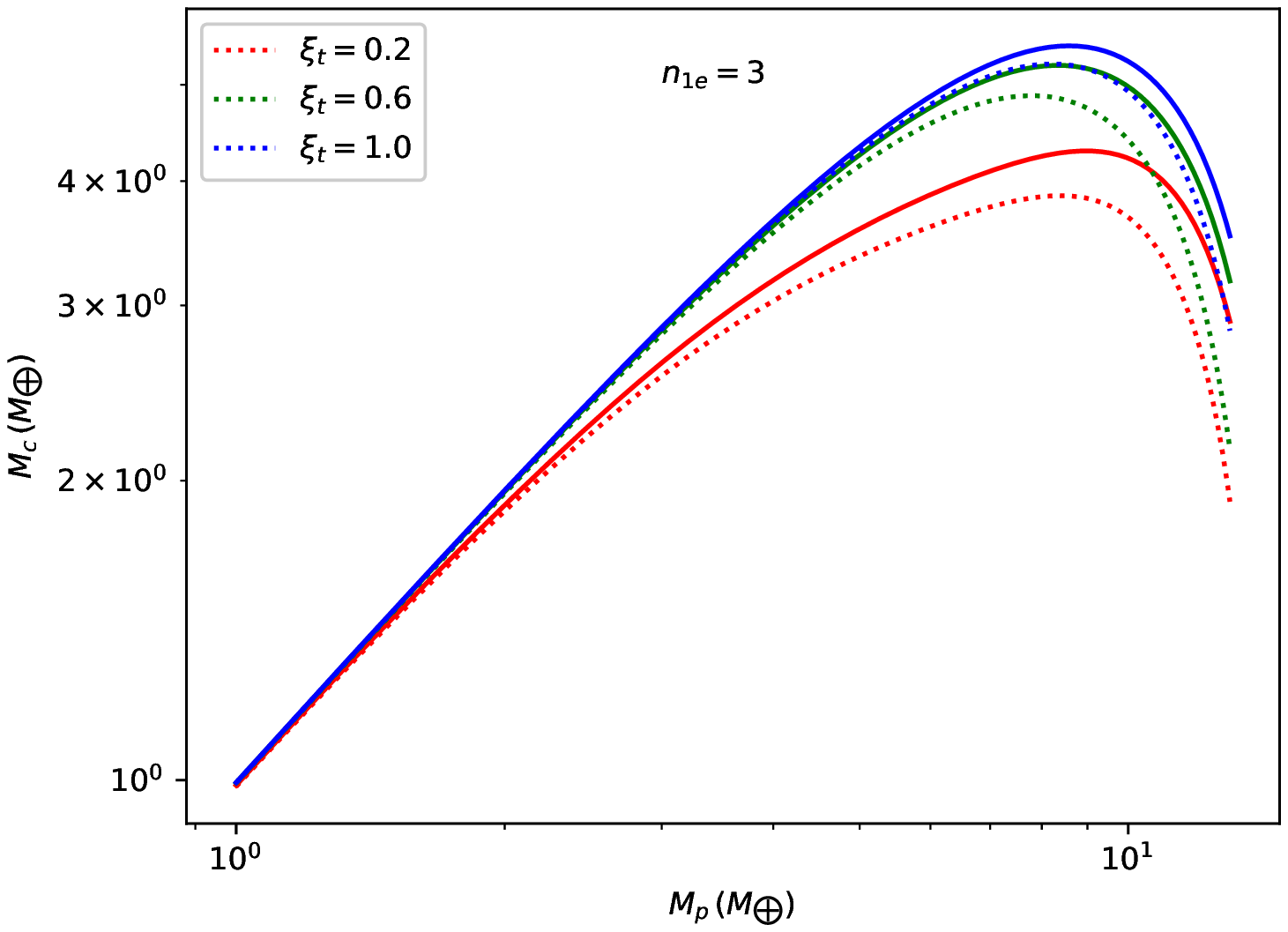}
\includegraphics[width = 8.5cm]{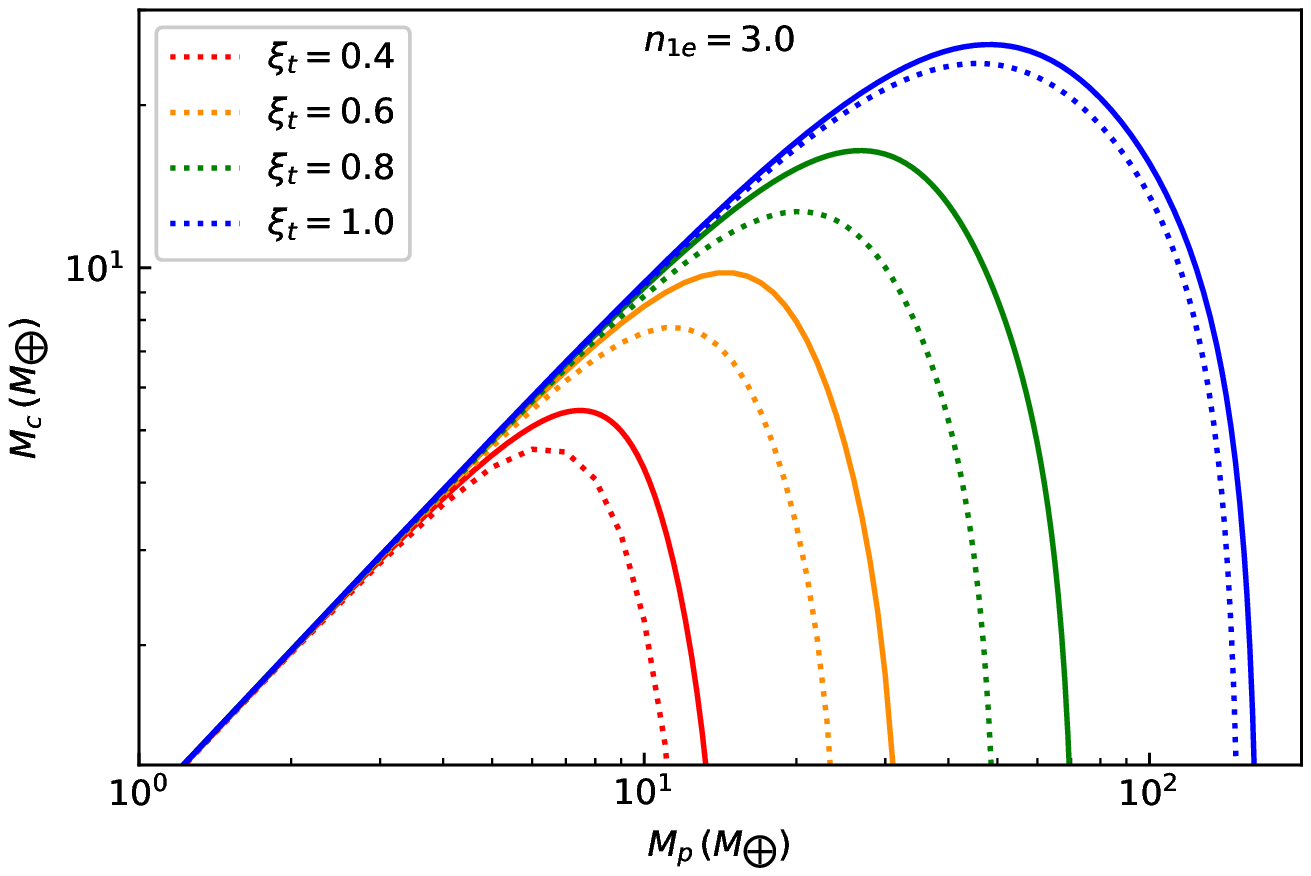}
    
\includegraphics[width = 8.5cm]{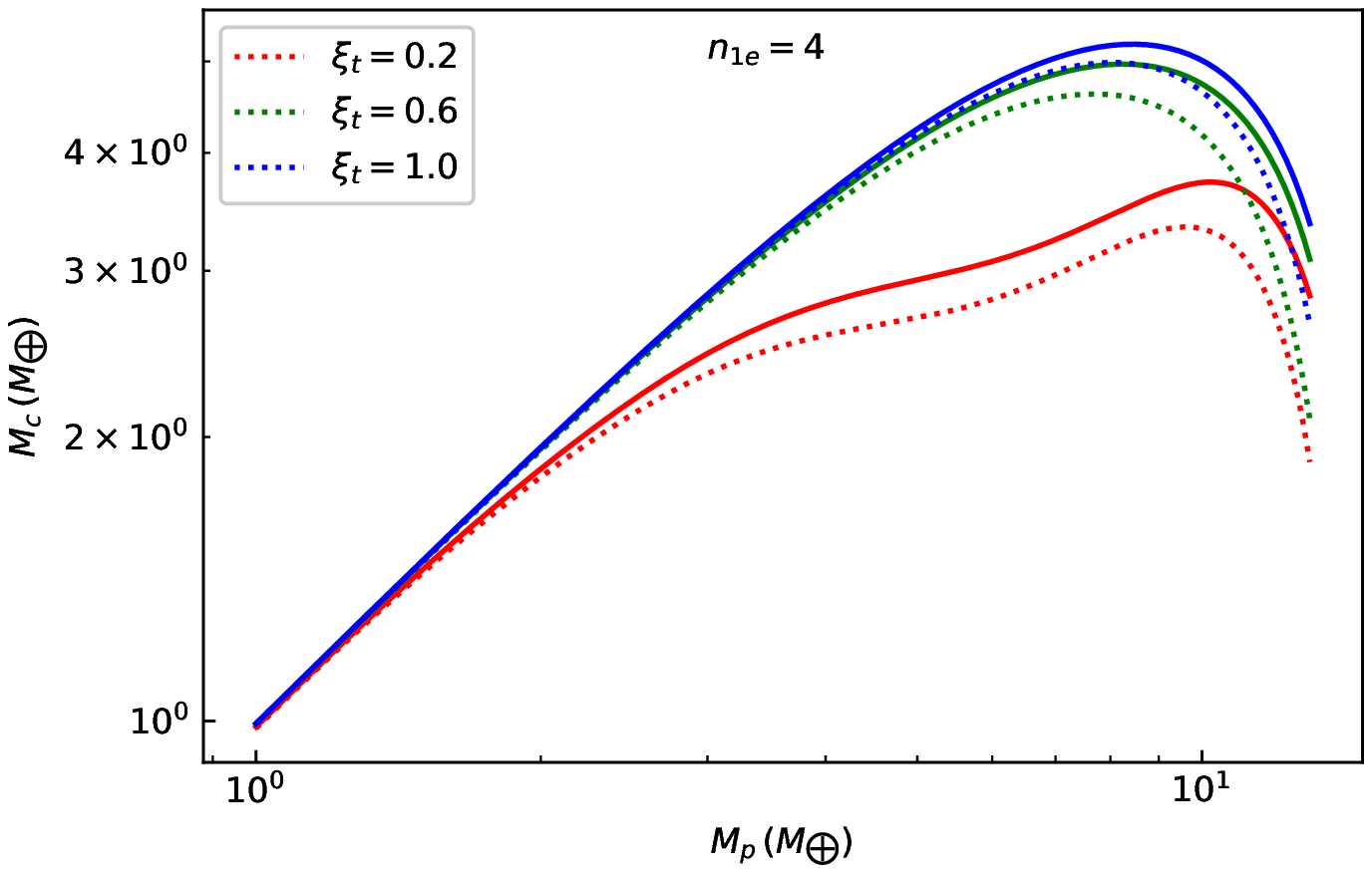}
\includegraphics[width = 8.5cm]{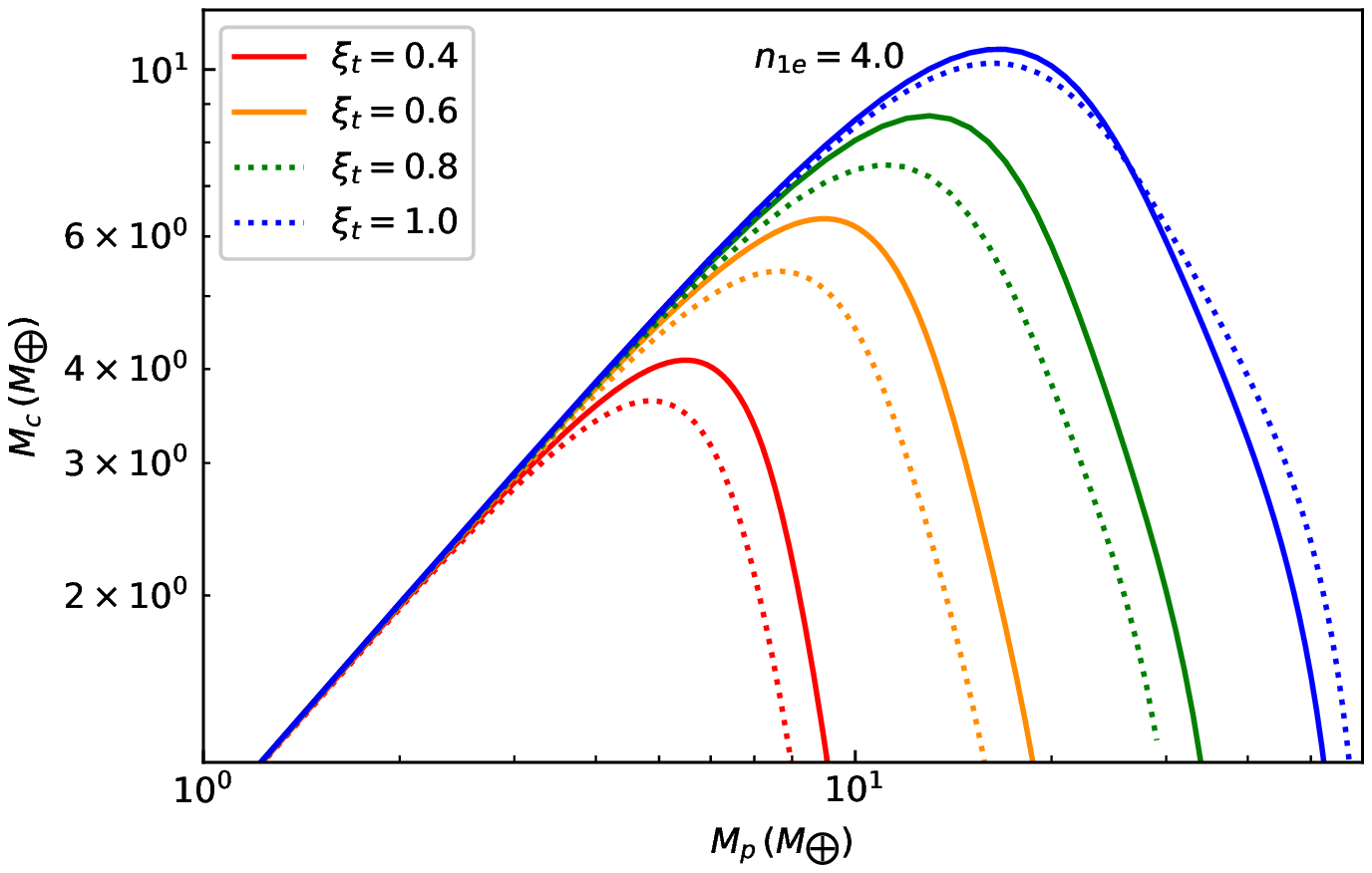}
\caption{The relationship of the composite polytopes between core mass and total mass. On the left is the Bondi model, and on the right is the Hill model. Different colors represent different radiation convection boundaries $\xi_{\rm t} = 0.2-1.0$.}
\label{fig:mcore_mp_composition}
\end{figure*}

In the Bondi models, $V_{\rm surf}f_{\rm p}/\left(n+1\right)<1$, the density distribution is the flattest and  is almost determined by the outer boundaries.  Rotation will increase the core gravity to keep the same total mass as the non-rotating case since the centrifugal force weakens the gravitational force. Subsequently, core mass increases. The maximum core mass is a transition point between the core gravity and the self-gravity of the envelope  \citep{2013ApJ...765...33K}, which can be derived when the surface structural line approaches the horizontal line
\begin{equation}
\begin{split}
U_{\rm surf}^{\rm crit} \simeq & \frac{\left(3\alpha+1\right)-\gamma_{\rm disk}f_{\rm p,surf}/\left(n+1\right)}{\alpha+1}\\
\simeq & \frac{3\alpha+1}{\alpha+1}-\frac{\gamma_{\rm disk}f_{\rm p,surf}}{\left(n+1\right)\left(\alpha+1\right)},
\end{split}
\label{U_crit}
\end{equation}
when $n=3$. The value of $U_{\rm surf}^{\rm crit}$ will increase due to $\alpha>0$ and $f_{\rm p,surf}<1$,  the corresponding total mass of the critical model will switch into
\begin{equation}
M_{\rm p}^{\rm crit} = \gamma_{\rm disk}^{3/2}\left[\frac{3\alpha+1}{\alpha+1}-\frac{\gamma_{\rm disk}f_{\rm p,surf}}{\left(n+1\right)\left(\alpha+1\right)}\right]^{\rm 1/2}M_{\rm 0},
\label{m_p_crit}
\end{equation}
combined with Equation (\ref{B_condition_u}).
The total mass of the non-rotating planet is nearly approximate to the characteristic mass $M_{\rm 0}$ \citep{2013ApJ...765...33K}. But the critical mass in a rotating planet will increase for  $\alpha>0$ and $f_{\rm p,surf}<1$. When $n<3$, the surface parameter $U_{\rm surf,B}$ is slightly larger than the $U_{\rm surf}^{\rm crit}$ so that the final mass of the critical model will increase. If $n>3$, 
$U_{\rm surf,B}$ is lower than  $U_{\rm surf}^{\rm crit}$, the total mass of the critical model will decrease.  Core mass can be confirmed by the jump condition. Combined the results in Table \ref{tab:critical}, the value of $V_{\rm 1e}^{\rm crit}$ for the critical model at the inner edge approaches $V_{\rm 1e}^{\rm crit}\sim \left(n+1\right)/f_{\rm p}$.  As the core mass can be increased by rotation, the variables at the inner edge may increase.

The structure lines for the rotating planets in the Hill model are shown on the right column of Figure \ref{fig:strucure_single}. In the soft polytrope ($n>3$), the inner edges are almost below the horizontal line. Unlike the Bondi model, the structure lines of the Hill model all across the horizontal, critical, and vertical lines. We analyze the characteristics of the structure line by the value of $V_{\rm surf}f_{\rm p}$. 

When $V_{\rm surf}f_{\rm p}/\left(n+1\right) \gg 1$, the structure line is above or is below the horizontal line. For the former, the atmospheric mass is less than core mass so that the core gravity determines the structural signatures. The structure line first increases $U$ and $V$ outward, in which the value of $V$ for the rotating planet is higher than that of the planet without rotation. Thus,  rotation will force the density to drop steeper, in the radial coordinate. Subsequently, the structure line decreases  $U$ outward. However, rotation in this state will reduce  $V$ and then form a flatter descent profile of the density. Thus, core density in a rotating planet is much higher. In the thinner envelope, the structural line near the outer boundary will spiral downwards, which will flatten the corresponding density distribution. For the latter, the structural line first decreases  $V$ outward. The value of $V $ for a rotating planet is higher than that of the non-rotating state, implying the density drops steeper. Once the structure line extends to the critical line, we can get the maximum $\left(U/V\right)_{\rm max}$. The structural line turns to increase  $V$, in which the signature is similar to the former. Most of the atmospheric  mass is concentrated in the middle of the planet, i.e., nearing the intersection with the critical line. Thus, we found that core mass will increase based on the complete density change trend.

In the case of $V_{\rm surf}f_{\rm p}> n+1$, core mass is determined by the innermost shell. The effect of rotation on the structure line is weaker compared to that of $V_{\rm surf}f_{\rm p}\gg n+1$. The tails of all structural lines starting at the inner edge will spiral downwards, and the rise in the corresponding mass becomes more gentle. The increase in the critical core mass are relatively slight. The critical models for $V_{\rm surf}f_{\rm p}\gg n+1$ and $V_{\rm surf}f_{\rm p}> n+1$ show the same signature. The value of  $\left(U/V\right)_{\rm max}$ for a rotating planet is comparable to that of the non-rotating state. The inner edge of the critical model nears the horizontal line, and  slope for the structural line of the critical model before reaching the critical line is almost zero, ${\rm d}\log V/{\rm d}\log U = 0$.  The values of  $V_{\rm 1e}^{\rm crit}$ and $U_{\rm 1e}^{\rm crit}$ for the critical model are listed in Table \ref{tab:critical}. The values of  $V_{\rm 1e}^{\rm crit} $ and $U_{\rm 1e}^{\rm crit}$ also increase, in which $V_{\rm 1e}^{\rm crit} \sim \left(n+1\right)/f_{\rm p}$.

In the case of $n=4$, $V_{\rm surf}f_{\rm p}/\left(n+1\right)$. The density distribution is relatively modest. When the atmospheric mass is below the core mass, core gravity dominates the evolution processes. Starting at the outer boundary, the slope of the structural line in a rotating planet is negative and is steeper. Compared with the non-rotating planet, the density has a more gentle incremental profile since rotation reduces $V$. In the positive slope, the density distribution becomes steeper since $V$ has increased by rotation. When the atmospheric mass grows to core mass, the structure line for a rotating planet (solid red line) on the right side of the surface boundary is higher than that of the non-rotating state. 
According to  \cite{2013ApJ...765...33K}, the inner edge of the critical model is just above the singular point and approaches
\begin{equation}
U_{\rm 1e}^{\rm crit} =\left \{\left [ 3\left ( \alpha +1\right )-2\right ] n -3\right \}/\left [ \left ( \alpha +1\right )n-1\right ].
\label{hill_condition}
\end{equation}
Rotation will increase the value of $U_{\rm 1e}^{\rm crit}$ due to $\alpha>0$.
The maximum core mass can be derived from Equation (\ref{tot_jump_condition}),  which can be shown as:
\begin{equation}
\begin{split}
M_{\rm core}^{\rm crit} \simeq & \left[\frac{\left [ 3\left ( \alpha +1\right )-2\right ] n -3}{\left ( \alpha +1\right )n-1}\right]^{3/2n}\left(\frac{n+1}{f_{\rm p}}\right)^{3/2}\\
& \times \left(\frac{3\rho_{\rm disk}}{\rho_{\rm core}}\right)^{\left(n-3\right)/2n}.
\end{split}
\label{critical_core_mass_hill}
\end{equation}
The critical core mass will increase for $f_{\rm p}<1$ and $\alpha>0$.
When the planet enters the runaway accretion stage, the structure lines will spiral downwards over the inner edge. Rotation increases the $V$ at the middle part, but the $V$ at both ends decreases. The inward decrease in mass undergoes the state of flatter, steeper, and flatter, respectively. In addition, the decrease in $V_{\rm 1e}$ is more  obvious than that of the non-rotating planet. The core mass of a rotating planet may be less than the non-rotating state under the runaway (see in the right panel of Figure \ref{fig:strucure_single}).

\subsection{The Results for The Composite Polytrope}\label{sec:result_isothermal}

In the previous section, we discussed the properties of the fully-convective planet (a single polytropic model).  Instead of a single structure, a two-layer structure with an exterior radiative layer and an interior convective layer is better suited to describe planet structure. \citep{2006ApJ...648..666R}.  In Section \ref{sec:core_mass_composite}, we study the effect of rotation on the core mass for the composite polytrope. In Section 3.2.2,  we research the influence of spin on the structural lines of the composite polytrope.

\subsubsection{the core mass of the composite polytrope}\label{sec:core_mass_composite}

\begin{figure*}
    \centering
    \includegraphics[width = 8.6cm]{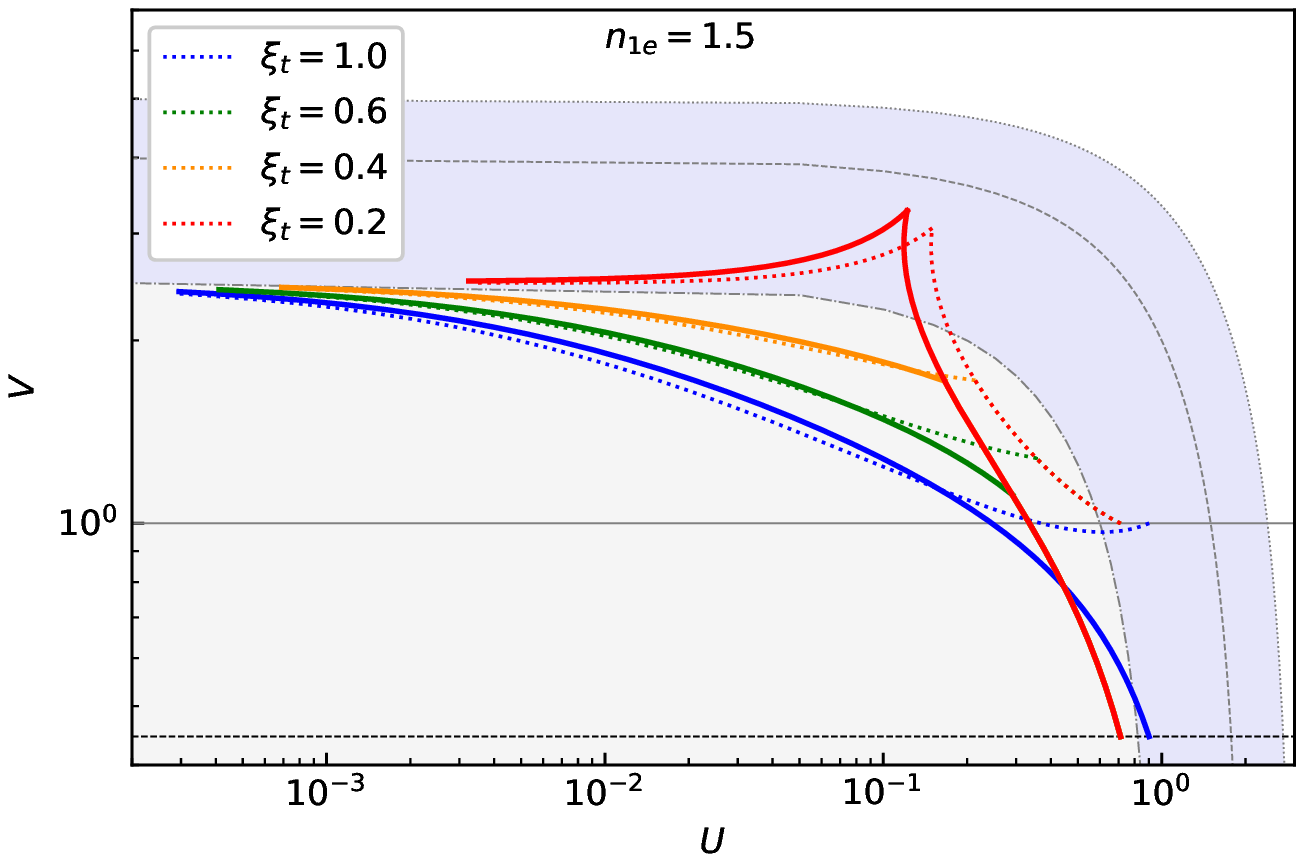}
    \includegraphics[width = 8.6cm]{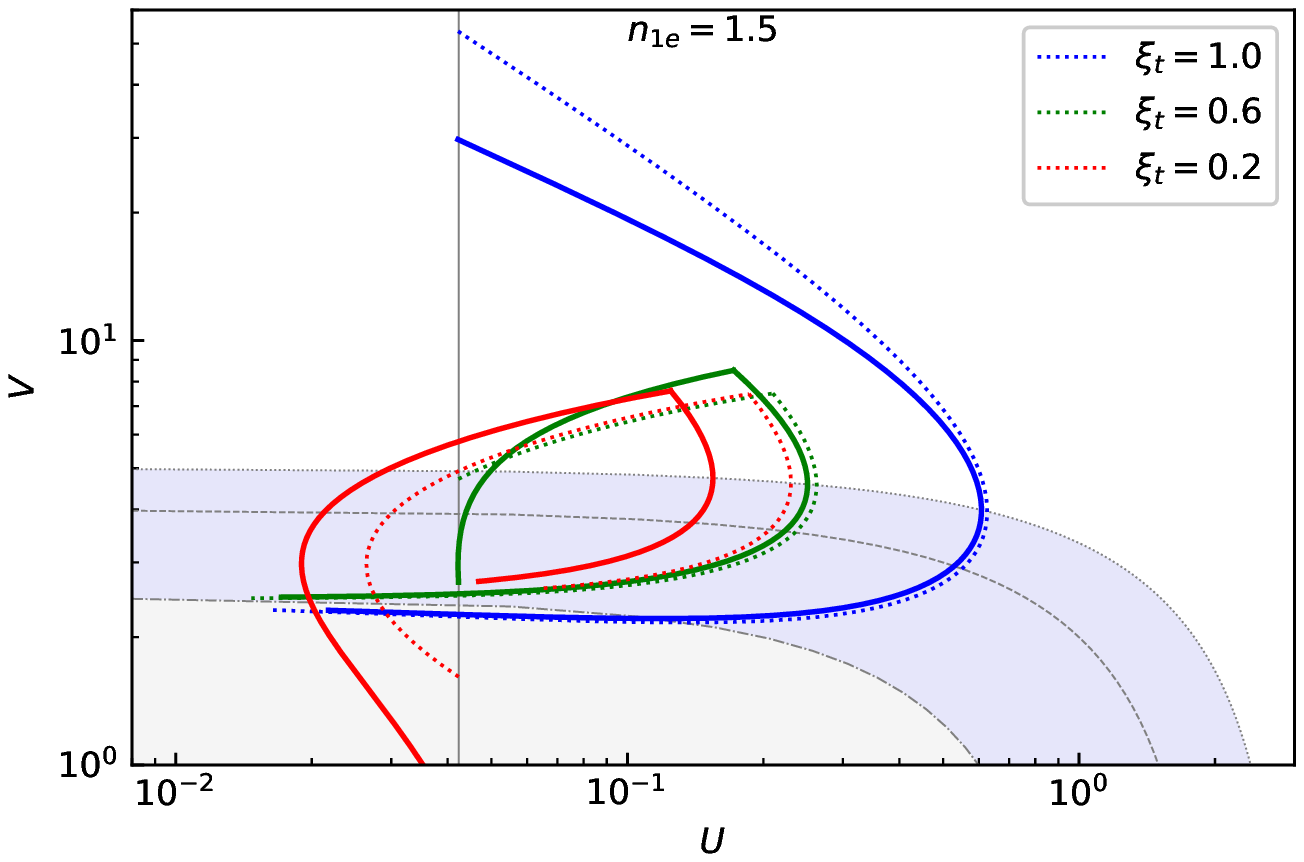}
    
    \includegraphics[width = 8.6cm]{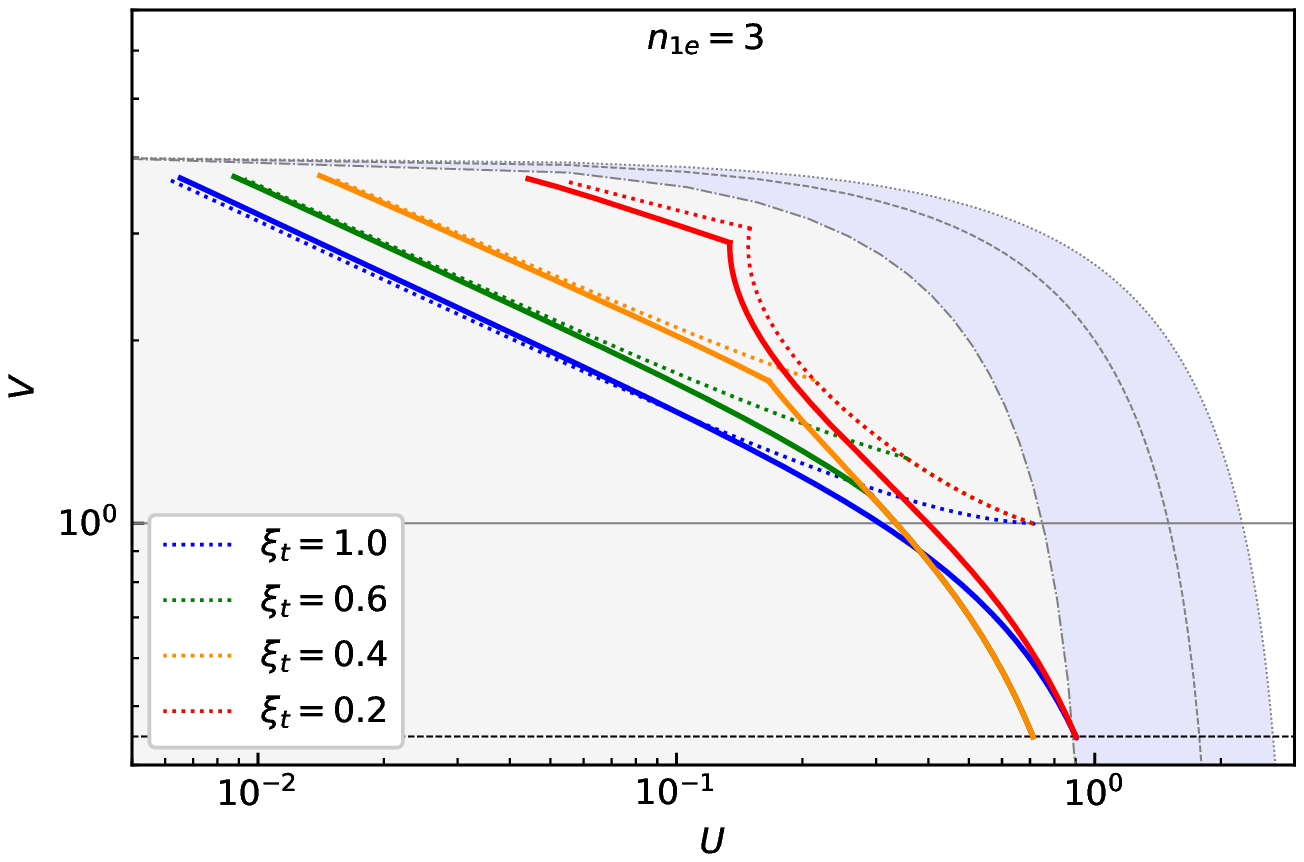}
    \includegraphics[width = 8.6cm]{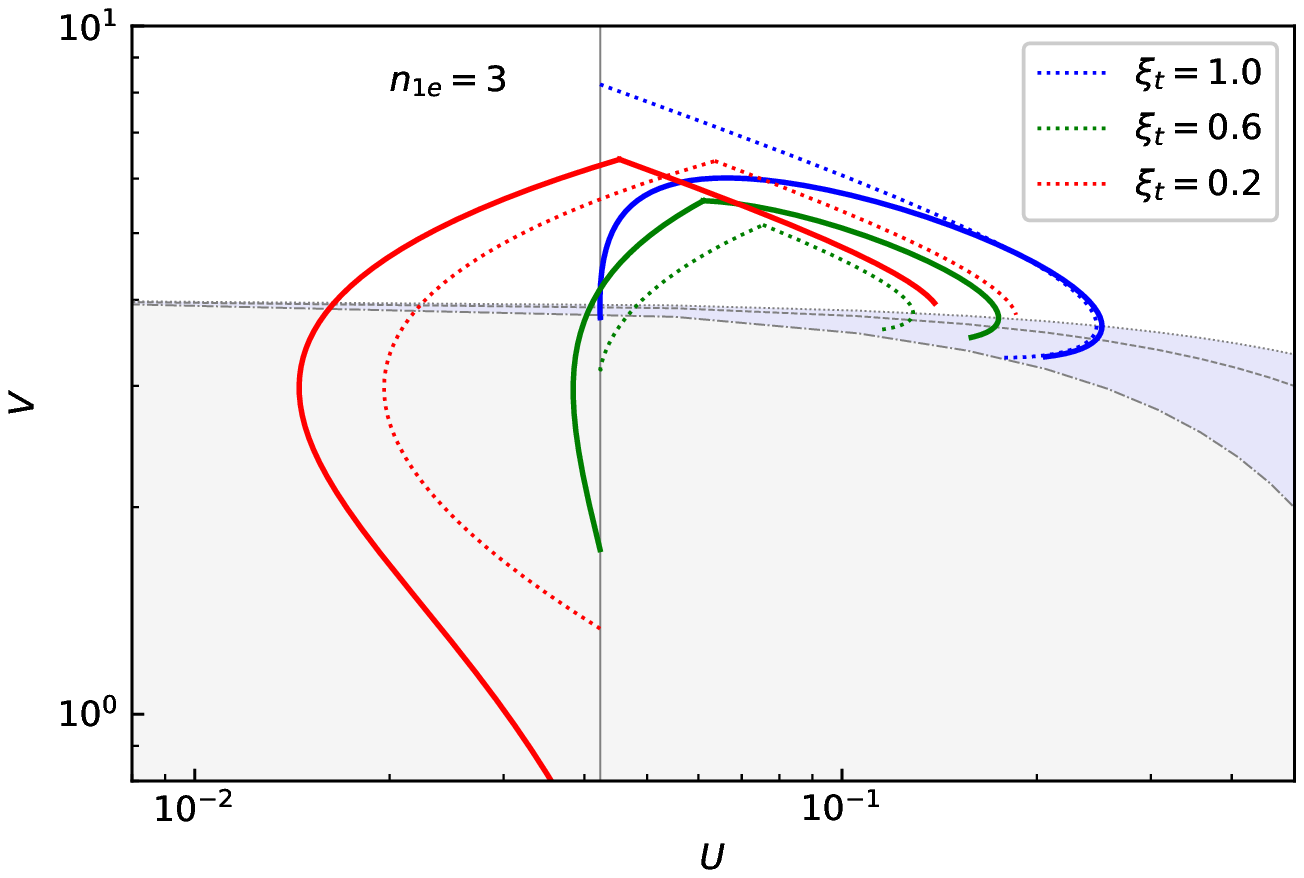}
    
     \includegraphics[width = 8.6cm]{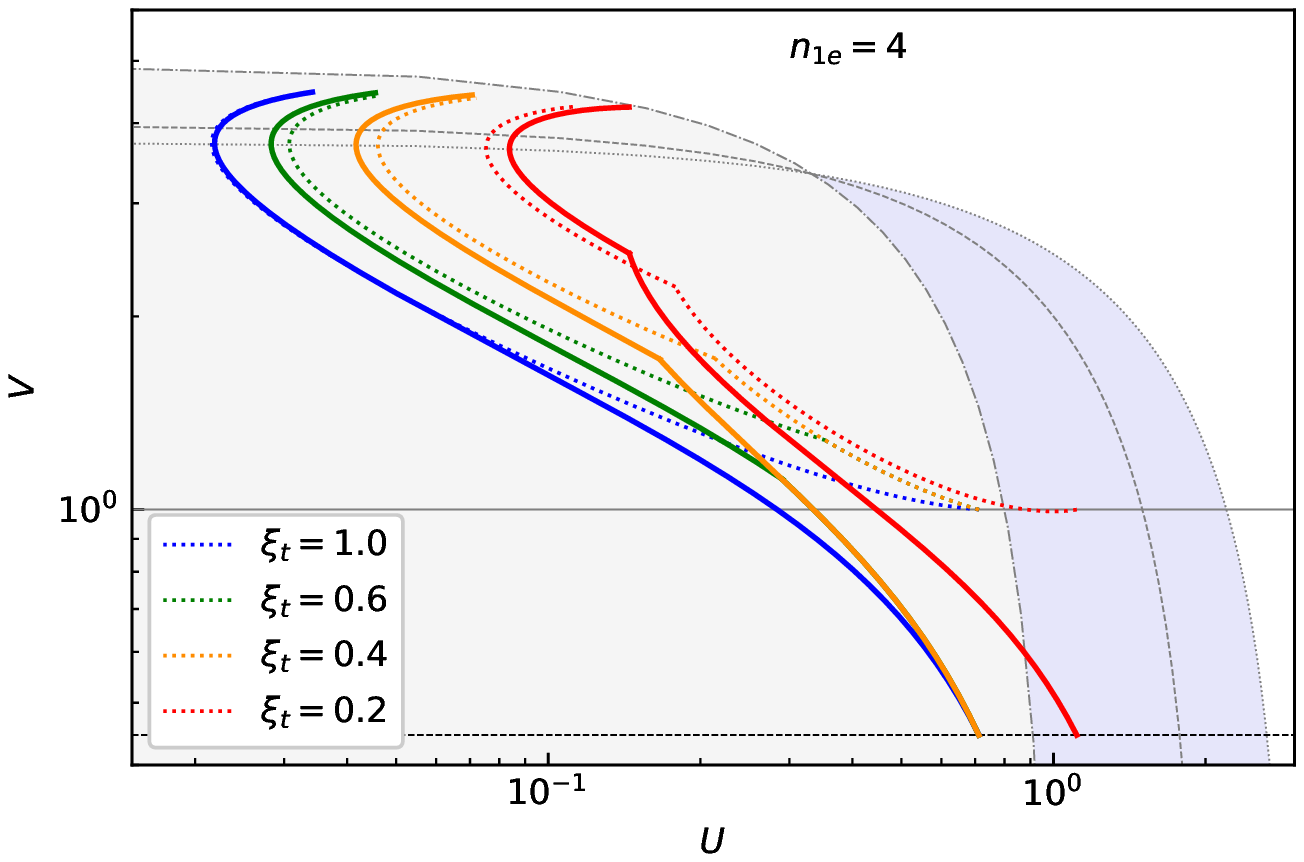}
    \includegraphics[width = 8.6cm]{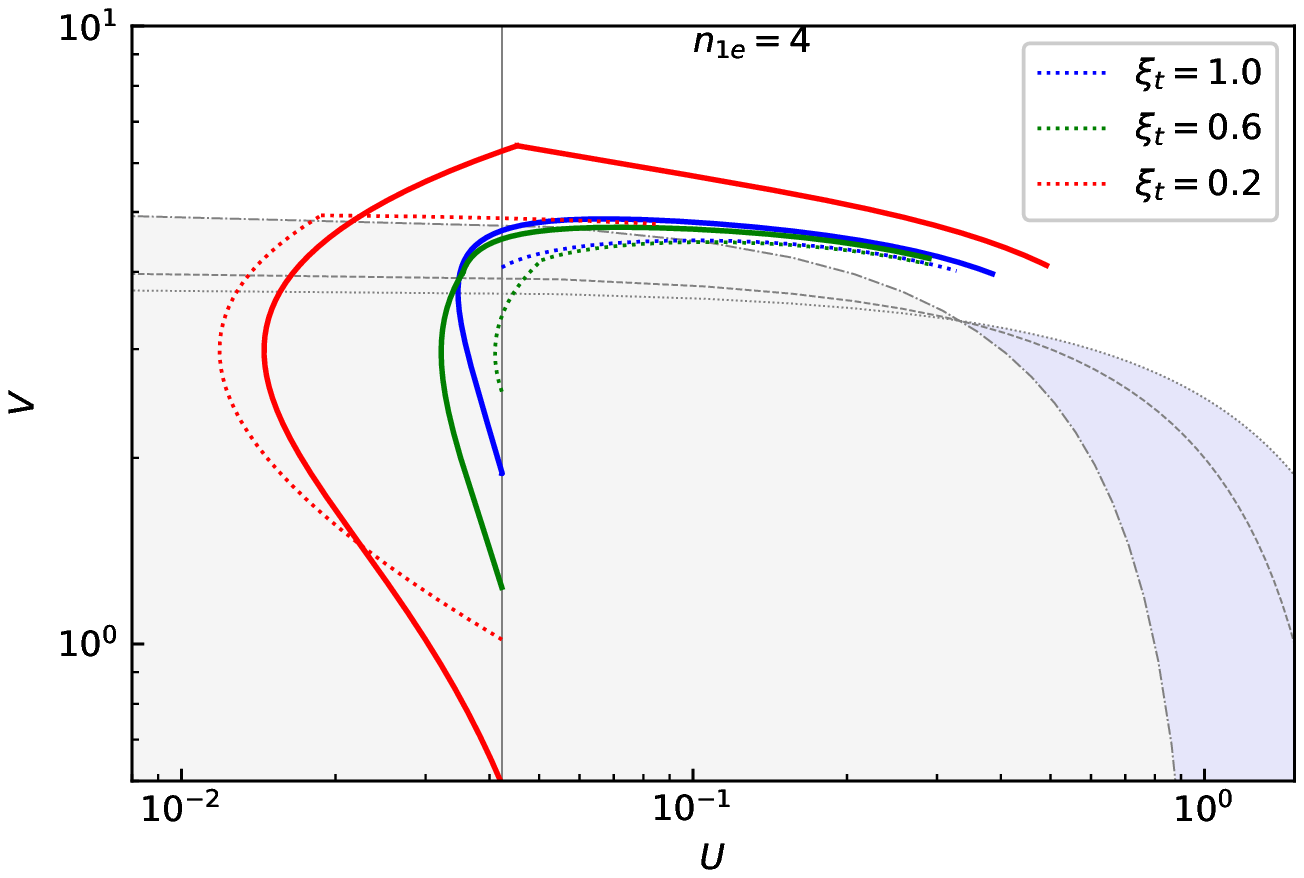}
    \caption{Behaviors on the characteristic plane for the critical model with the different thicknesses of isothermal layer. Different colors correspond to the different thicknesses. The left panels represent the Bondi models, and the right panels denote the Hill models. Solid and dotted lines correspond to the rotating and the non-rotating planets, respectively.
}
    \label{fig:the_isothermal_strucutre}
\end{figure*}

The planetary shape may be a composite polytrope  \citep{2013ApJ...765...33K}, in which the interior convective layer may be covered with a radiative layer \citep{2006ApJ...648..666R}. The polytropic index of the radiative layer approaches  $n=\infty$, but $n_{\rm 1e} = 1-5$ for the convective layer. There is a transition radius, $R_{\rm t} = \xi_{\rm t} R_{\rm out}$ (the transition radius ratio, $\xi_{\rm t} = 0.1-1 $), at the radiative-convective boundary.

The radiative layer can be treated as the isothermal layer, the temperature in this layer seems to be  $T \sim T_{\rm disk}$ \citep{2014ApJ...797...95L,2015ApJ...811...41L}.  The polytropic equation of state in the radiative layer can be read as  \citep{2013Stellar}
\begin{equation}
P = K_{\rm disk} \rho.
\label{eq:radiative_pressure}
\end{equation}
 The polytropic constant for the radiative envelope is $K_{\rm disk}=P_{\rm disk}/\rho_{\rm disk}$. Thus, we will derive the density at the transition radius (i.e., the radiative-convective boundary, RCB) by Equation (\ref{V})
\begin{equation}
\begin{split}
\rho_{\rm t} = & \rho_{\rm disk}\exp{\left[\frac{GM_{\rm core}\rho_{\rm disk}f_{\rm p}}{p_{\rm disk}R_{\rm out}}\left(\frac{1}{\xi_{\rm t}}-1\right)\right]}\\
=& \rho_{\rm disk}\exp{\left[V_{\rm surf}f_{\rm p}\left(\frac{1}{\xi_{\rm t}}-1\right)\right]}.
\end{split}
\label{eq:rho_rcb}
\end{equation}
In the non-rotating case ($f_{\rm p}=1$),  when the value of  $\xi_{\rm t}$ decreases, $\rho_{\rm t}$ becomes larger, and the density in the radiative layer increases exponentially.  The variable $\rho_{\rm t}$ is a new outer boundary for the convective layer, where the polytropic constant of the convective layer, $K_{\rm t}=K_{\rm disk}/\rho_{\rm t}^{\rm 1/n_{\rm 1e}}=P_{\rm disk}/\rho_{\rm disk}\rho_{\rm t}^{\rm 1/n_{\rm 1e}}$. The influence of the gravity of the envelope is enhanced since a smaller $K_{\rm t}$ requires a larger density to achieve hydrostatic equilibrium. Thus, the density distribution in the convection layer can be modified to
\begin{equation}
\begin{split}
\rho_{\rm gas}\left(r\right)= &\rho _{\rm t}\left [ 1+\frac{GM_{\rm core}\rho_{\rm disk}f_{\rm p}}{p_{\rm disk}\left ( n_{\rm 1e}+1\right )}\left ( \frac{1}{r}-\frac{1}{\xi_{\rm t}R_{\rm out} }\right )\right ]^{\rm n_{\rm 1e}}\\
=&\rho_{\rm t}\left[1+\frac{V_{\rm surf}f_{\rm p}}{n_{\rm 1e}+1}\left(\frac{R_{\rm out}}{r}-\frac{1}{\xi_{\rm t}}\right)\right]^{\rm n_{\rm 1e}}.
\end{split}
\label{eq:rho_conv}
\end{equation}
The values of $f_{\rm p}$ and $\xi_{\rm t}$ determine the convective density distribution. To obtain the core mass for a rotating case, we can employ Equations (\ref{eq:rho_rcb}) and (\ref{eq:rho_conv}) to analyze the features of the composite polytrope. When $M_{\rm core}$, $R_{\rm out}$, and $\xi_{\rm t}$ are constant, the density at the transition shell $\rho_{\rm t}$ falls significantly along with $f_{\rm p}$, resulting in a significant reduction in $\rho_{\rm gas}$, the density at the convective layer. Thus, the value of $M_{\rm core}$ in a rotating state is far lower than in the non-rotating situation, which is the opposite of the initial condition. The same $M_{\rm core}$ can be obtained by reducing the total mass or the outer radius. Under these conditions, a much higher core gravity (core mass) is required for the same outer boundary conditions. Based on the analysis above, the core mass becomes larger as $f_{\rm p}$ reduces. 

As seen in Figure  \ref{fig:mcore_mp_composition}, the evolution of the core mass with the total mass in the Bondi and Hill models are shown in the left and right panels, respectively. \cite {2013ApJ...765...33K} have verified the total mass (i.e., $M_{\rm core}$) in the critical model decreases with the transition radius ratio.  However, core mass can be increased by the rotation and shows a different signature with different thicknesses of the isothermal layer. Note that the isothermal layer will be removed when $\xi_{\rm t}=1.0$. For the rotating Bondi models, the break-up speed is proportional to the total mass due to $\omega _{\rm crit} = c_{\rm s}^{\rm 3}/GM_{\rm p}$. Thus, the angular velocity increases as the  $\xi_{\rm t}$ decreases. In the smaller envelope, the total mass is much less than that of the core, the parameter $\eta = \omega^2 r^3/2GM_{\rm r} \approx \omega^2 r^3/2GM_{\rm core}$. If the core mass and radius are the same as the non-rotating planet, $\eta$ will increase as the isothermal layer grows thick.  The coefficient of centrifugal force, $f_{\rm p}=3-2\left(1-\eta\right)^{\rm -2/3}$, under this situation will decrease sharply. The convective density will decrease sharply according to Equations  (\ref{eq:rho_rcb}) and (\ref{eq:rho_conv}). Following the analysis above, core mass increases more sharply when the isothermal layer becomes sufficiently thick.

In the Hill model, the break-up speed for a rotating planet is proportional to the sum of the mass of the central star and the planet and can be expressed as $\omega_{\rm crit} \propto \sqrt{\left(M_{\star}+M_{\rm p} \right)}$. Since the total mass of the critical model increases with the decrease in the thickness of the isothermal layer, the angular velocity $\omega=0.96\,\omega_{\rm crit}$ will increase. When the isothermal layer is thinner, we assume the same core mass, outer radius, and total mass as the non-rotating planet. The variable $\eta = \omega^2 r^3/2GM_{\rm r} \approx \omega^2 r^3/2GM_{\rm core}$  at the same radius will increase, resulting in a significant reduction in the coefficient of centrifugal force,  $f_{\rm p}=3-2\left(1-\eta\right)^{\rm -2/3}$. The convective density and core mass will drop noticeably,  which is the opposite of the assumption. To obtain the same core mass, we can increase core gravity by decreasing the outer radius or the total mass. When the planet's mass is given, a higher core gravity is required to keep the same mass as the non-rotating planet. In summary, the increase in the core mass is even sharper as the isothermal layer becomes thinner. In addition, the signature become more obvious as the value of $n_{\rm 1e}$ decreases.

\subsubsection{the structure of the composite polytrope}\label{sec:critcal_composite}

The  exterior  isothermal  layer  enables  the  structure line to cross  the horizontal  and critical  lines since  the inward decrease  in  its  thermal  energy  \citep{2013ApJ...765...33K}. According to Equation (\ref{V}), we can derive the homologous relation at the transition radius
\begin{equation}
\log \left(\frac{V_{\rm t}}{V_{\rm out}}\right) = \int_{\xi}^{1}(1-U)d\log \xi_{\rm t} + \log \left(\frac{f_{\rm p,t}}{f_{\rm p,out}}\right),
\label{eq:logV}
\end{equation}
with the characteristic variable at the surface $V_{\rm out}$, the coefficient of centrifugal force at the surface $f_{\rm p,surf}$ and transition radius $f_{\rm p,t }$. The coefficient of centrifugal force increases  inward and causes a sharp drop of $V_{\rm out}=GM_{\rm p}\rho_{\rm disk}{f_{\rm p,out}}/R_{\rm out}P_{\rm disk}$. 

The structural line of the critical model with the Bondi boundary is shown in the right column of Figure \ref{fig:the_isothermal_strucutre}. If $V_{\rm t}$  is under the horizontal line, the structure line will keep the same features as a single polytrope.  Rotation can decrease the radiative $V$ and steepen the radiative structural line, and then, the density has a more gentle descent profile. However, the slope of the convective structural line is almost similar to the non-rotating state. Thus, the change in the density at the transition shell will determine the core gravity.  When the radiative layer grows sufficiently thick, $V_{\rm t}$ is above the horizontal line, and then, the convective structural line will decrease inward. 

We can derive the condition of the core mass by the jump condition 
\begin{equation}
\begin{split}
V_{\rm 1e}U_{\rm 1e}^{\rm 1/n_{\rm 1e}} = &\frac{GM_{\rm core}}{R_{\rm core}}\frac{f_{\rm p}}{K_{\rm t}}\left(\frac{3}{\rho_{\rm core}}\right)^{\rm 1/n_{\rm 1e}}\\
= & \left(\frac{M_{\rm core}}{M_{\rm 0}}\right)^{\rm 2/3}\left(\frac { \rho_{\rm core}}{3\rho_{\rm disk}}\right)^{\rm \left(n_{\rm 1e}-3\right)/3n}  \left(\frac{ \rho_{\rm t}}{ \rho_{\rm disk}}\right)^{1/n_{\rm 1e}} f_{\rm p}.
\end{split}
\label{eq:jump_condition_new}
\end{equation}
Following \cite{2013ApJ...765...33K}, core mass decreases as the thickness of the isothermal layer increases.  When the effect of rotation is included, the condition of core mass can be reads as 
\begin{equation}
\begin{split}
M_{\rm core} = &\left(\frac{3\rho_{\rm disk}}{\rho_{\rm core}}\right)^{\rm \left(n_{\rm 1e}-3\right)/2n_{\rm 1e}} \left( \frac{1}{\rho_{\rm t}^{1/{\rm n_{\rm 1e}}}f_{\rm p}}\right)^{\rm 2/3}\\
& \times \left(V_{\rm 1e}U_{\rm 1e}^{\rm 1/n_{\rm 1e}}\right)^{\rm 2/3}\rho_{\rm disk}^{\rm 2/3n_{\rm 1e}} M_{\rm 0},
\end{split}
\label{eq:core_mass_rcb}
\end{equation}
with the given values of $U_{\rm 1e}$ and $V_{\rm 1e}$. When $n_{\rm 1e}>3$,  $\left(n_{\rm 1e}-3\right)/2n_{\rm 1e}<0$  and then the core mass will increase mildly with the core density. In addition, when $n_{\rm 1e}\leq 3$, the effect of the core density is slight.  Thus, the effect of the core density can be ignore. We note that the core mass, which has dropped by the isothermal layer, will be increased by the decrease in $f_{\rm p}$.

As seen in the left column of Figure 4, $V_{\rm 1e}$ increases slightly, compared with the non-rotating planet.  In addition, the $U_{\rm 1e}$ changes in different manners have a mild effect on the core mass. Thus, core mass in the Bondi model mainly determines by the polytropic constant $K_{\rm t}$ according to Equation (\ref{eq:core_mass_rcb}). As mentioned above, the coefficient of centrifugal force $f_{\rm p}$ in the Bondi model decreases with the $\xi_{\rm t}$. It will force the density at the transition shell to drop modestly and then increase $K_{\rm t}$. 
As a result, the increase in the core mass and the total mass of the critical model are more significant when the isothermal layer grows sufficiently  thick. In addition, the total mass of the critical model in the rotating planet will increase since the parameter $\alpha $ increases with the decrease in $f_{\rm p}$ according to Equation (\ref{m_p_crit}).

The structural line of the critical model with the Hill boundary  is shown in the left column of Figure \ref{fig:the_isothermal_strucutre}. The structure line of a composite polytrope is different from a single polytrope. The outer boundary  ($V_{\rm out}=V_{\rm surf,H}$) reduces sharply by the coefficient of centrifugal force. In addition, the ratio of the characteristic variable $V_{\rm out}/V_{\rm t}$ reduces with $\xi_{\rm t}$ \citep{2013ApJ...765...33K}, which will force the isothermal layer to spiral downward.  The slope of the isothermal layer satisfies
\begin{equation}
\frac{{\rm d}\log V}{{\rm d}\log U} \simeq \frac{1-{\rm d}\log f_{\rm p}/{\rm d}\log r}{V-3},
\label{eq:isothermal_slope}
\end{equation} 
with ${\rm d}\log f_{\rm p}/{\rm d}\log r < 0$ and $U\ll 1$.   When the isothermal layer grows sufficiently thick, the structure will spiral downward, and $V_{\rm out}$ is below the horizontal line. As  $U$ decreases inward,  the structure line becomes negative and steepens by rotation, forcing the mass distribution to decrease gently. However, the changes in the structure line with a positive slope show the opposite manner of the former. Since $f_{\rm p}$ increases inward, the slope will gradually become similar to the non-rotating planet.   $V_{\rm t}$ is above the critical line. 
Once $V$ reaches $V_{\rm t}$, the structure line has turned, and then, the slope of the adiabatic lines in the stiff polytrope becomes negative due to
\begin{equation}
\frac{{\rm d}\log V}{{\rm d}\log U} \simeq - \frac{1+\left({\rm d}\log f_{\rm p}/{\rm d}\log r\right)\left(n_{\rm 1e}+1\right)/V}{n_{\rm 1e}},
\end{equation}
with $V \gg n_{\rm 1e}+1$.  The value of  $\left({\rm d}\log f_{\rm p}/{\rm d}\log r\right)\left(n_{\rm 1e}+1\right)/V$ is tiny as $V \gg n_{\rm 1e}+1$, resulting in a flatter slope and a mild change in the thinner convective layer.  The mass distribution drops steeper as the structural line becomes flatter. Thus, we can get the complete trends of the mass and density distribution. 

The condition of the core mass for the Hill model is mainly determined by the $V_{\rm 1e}$, $U_{\rm 1e}$, and $K_{\rm t}$ following Equation  (\ref{eq:core_mass_rcb}).  As mentioned above, $f_{\rm p}$ in the Hill model will decrease as the isothermal layer is thinner.  Compared with the non-rotating planet, rotation under this situation will decrease $\rho_{\rm t}$ more significant, forcing a sharper increase in $K_{\rm t}$. 
For $n_{\rm 1e}\leqslant  3$, the increase in  $V_{\rm 1e}$ is mild. As the isothermal layer is thinner, $U_{\rm 1e}$ in a rotating planet becomes greater.  However, rotation decreases $U_{\rm 1e}$ when the isothermal layer grows thick. Thus,  rotation will noticeably increase the core mass when the isothermal layer becomes thinner. For $n_{\rm 1e}>3$ and the isothermal layer is thinner, rotation mildly increases $U_{\rm 1e}$. However, there are a sharper increase in $U_{\rm 1e}$ and a slight decrease in $V_{\rm 1e}$ for a rotating planet when the isothermal layer is thicker. Combined with the polytropic constant $K_{\rm t}$, the increase in the core mass for the Hill model is significant as the isothermal layer is thinner, especially for the stiff Hill model. In a word, core mass increase with the decrease in the thickness of the isothermal layer and the polytropic index.

 In short, the isothermal layer is a cooling shell, which can reduce the thermal energy or the polytropic index \citep{2013ApJ...765...33K}. Thus, the core mass will be decreased. However, rotation in the composite polytrope will noticeably increase the core mass compared with the non-rotating state. It will change the surface parameters and the signature of the structure line. Although the change in the Bondi model is slight, a significant change will be induced in the structure line of the Hill model by rotation with the isothermal layer.

\section{conclusion}\label{conclusion}

Rotation can deform the shape of the planet. The centrifugal forces, driven by spin, would change the hydrostatic equilibrium. Thus, rotation will alter the critical core mass and determine whether the planet triggers runaway accretion.

Since rotation can particularly weaken the gravitational force, core gravity will enhance to keep the same mass as the non-rotating planet, forcing an increase in core mass. The critical core mass is determined by the polytropic index, outer radius, or isothermal layer. Core mass in the Bondi and the soft Hill model will slightly increase by rotation. However, the increase in core mass for the stiffest Hill model is more significant with higher core gravity. Rotation can also enhance the critical core mass conditions according to the homology relationship. In addition, the existence of the isothermal layer will increase the planetary density and then decreases the critical core mass. However, planetary rotation can reduce this density by centrifugal force, and new conditions are established for the convective  density. The critical core mass then increases considerably with reductions in $\rho_{\rm t}$ and $f_{\rm p}$. Since the angular velocity in the Hill model is proportional to the planetary mass, the critical core mass will increase sharply when the radiative layer becomes thinner. However, the case with the Bondi model is the opposite. In short, rotation slightly increases the critical core mass of a single polytrope. It has a tremendous effect on the growth of critical core mass in the stiff Hill  and a composite polytrope. 

The structure and formation of the exoplanet may be deeply effected by rotation.  However, in recent years there have been few studies on how spin affects the formation of planets. This study considers a barotropic state, where the planetary shape is deformed. We have verified the effect of the isothermal layer of a rotating planet on the critical core mass. Rotation will increase the critical core mass and then prolong the evolutionary timescale (i.e., KH contraction timescale), which will inhibit the runaway accretion or the formation of gas giant planets. Subsequently, Mini-Neptunes/Super-Earths may develop in the post-formation stages.

There are some limitations to this work. Firstly, we assumed that an isothermal layer exists with an infinite polytropic index. The isothermal layer will add a new temperature gradient to the planet. However, the trend of this temperature gradient is compatible with cases without rotation. We did not consider the coefficient $f_{\rm R}>1$ \citep{2002A&A...394..965Z} in the radiative temperature gradient, which will lead to changes in the polytropic relationship between pressure and density. Secondly, the transition radius (i.e., radiative-convective boundary) is fixed in the rotating planets. When the transition radius of the polytropic model evolves with time, the critical core mass and the planetary formation may show different signatures.

In general, rotation is a two-dimensional \citep{2019MNRAS.487.2319B}  or three-dimensional problem \citep{2019MNRAS.488.2365B}. When the core mass is greater than the thermal mass, the envelope gets considerable rotation support  \citep{2019MNRAS.487.2319B}. Meanwhile, the one-dimensional rotation model is no longer applicable. However, we can introduce the vortex factor to a one-dimensional model to study the effect of the vortex with rotation support on the structure and evolution of the planet. Spin in this state may have a significant influence on planet formation. This work bases on a simple core accretion model. However, \cite{2018MNRAS.479..635K} proposed the atmospheric recycling is a complex two-dimensional accretion model, which can slow or stall envelope accretion. Thus, atmospheric recycling also can change the accretion rate and the critical core mass. In addition, rotating planets may also be affected by the magnetic dynamo \citep{2019MNRAS.488.5633K} in which the gravitational harmonics would be compatible with the actual data from Junno or Cassini spacecraft. Rotation works differently under different boundary conditions, so if we consider the evolution of the disk, it may lead to different results.

\section{acknowledgments}
We thank the anonymous referee for his/her suggestions that greatly improved this paper. 
This work has been supported by National Key R \& D Program of China (No. 2020YFC2201200), the science research grants from the China Manned Space Project (No. CMS-CSST-2021-B09 \& CMS-CSST-2021-A10), and opening fund of State Key Laboratory of Lunar and Planetary Sciences (Macau University of Science and Technology) (Macau FDCT grant No. 119/2017/A3).  C.Y. has been supported by the National Natural Science Foundation of China (grants 11373064, 11521303, 11733010, and 11873103), Yunnan National Science Foundation (grant 2014HB048), and Yunnan Province (2017HC018).

\vspace{5mm}



\bibliography{sample631}{}
\bibliographystyle{aasjournal}



\end{document}